\documentclass[preprint,12pt]{elsarticle}

\usepackage{graphics}
\usepackage{amssymb}
\usepackage{rotating}
\usepackage{xcolor}

\usepackage{colortbl}
\usepackage{color}
\usepackage{colordvi}
\journal{\textcolor[rgb]{0.93,0.93,0.93}{vv}}

\begin{document}

\nocite{*}
\begin{frontmatter}



\title{A comparative simulation study of data-driven methods for estimating density level sets}


\author[rvt]{P. Saavedra-Nieves \corref{cor1}}
\author[rvt]{W. Gonz\'alez-Manteiga}
\author[rvt]{A. Rodr\'iguez-Casal}
\address[rvt]{Department of Statistics and Operations Research, University of Santiago de Compostela, Spain}
\cortext[cor1]{Corresponding author: paula.saavedra@usc.es (P. Saavedra-Nieves)}

%\begin{abstract}
%The problem of density level set estimation has received considerable attention in the literature. From a random sample generated by the corresponding density, %the level set which is defined as the set of points where the density is above a given threshold, can be reconstructed using three methodologies: Plug-in, excess %mass and hybrid methods. The first one is based on replacing the unknown density by some nonparametric estimator, usually the kernel one. So, the bandwidth %selection problem is fundamental from the practical point of view. Classical methods for selecting the bandwidth are designed to reconstruct the density function. %Specific selectors for level sets have been proposed recently. In the other hand, if some information a priori about the shape of the level set is available then %excess mass algorithms can be useful. In this case, a nonparametric density estimator is not necessary and the problem of bandwidth selection can be avoided. The %last methodology is a hybrid of the two previous ones. Just as the excess mass, it assumes a geometric restriction on the level set and, like the plug-in methods, %it needs a pilot nonparametric estimator of the density. One interesting open question is about the practical performance of these methods. In this work existing %methods are reviewed and two new hybrid algorithms are proposed. Their practical behavior is compared through an extensive simulation study. Finally, some useful %conclusions for practitioners are presented.

%\end{abstract}

\begin{abstract}
Density level sets are mainly estimated using one of three methodologies: plug-in, excess mass, or a hybrid approach. The plug-in methods are based on replacing the unknown density by some nonparametric estimator, usually the kernel. Thus, the bandwidth selection is a fundamental problem from a practical point of view. Recently, specific selectors for level sets have been proposed. However, if some a priori information about the geometry of the level set is available, then excess mass algorithms can be useful. In this case, a density estimator is not necessary, and the problem of bandwidth selection can be avoided. The third methodology is a hybrid of the others. As in the excess mass method, it assumes a mild geometric restriction on the level set and, like the plug-in approach, requires a pilot nonparametric estimator of the density. One interesting open question concerns the practical performance of these methods. In this work, existing methods are reviewed, and two new hybrid algorithms are proposed. Their practical behaviour is compared through extensive simulations.

\end{abstract}


\end{frontmatter}

% \linenumbers


\section{Introduction}\label{introduction}

\normalsize{
Level set estimation theory deals with the problem of reconstructing an unknown set $L(\tau)=\{f \geq f_\tau\}$ from a random sample of points $\mathcal{X}_n=\{X_1,...,X_n\}$ of a random variable $X$, where $f$ is the density of $X$, $\tau\in (0,1)$ is a given probability, and $f_\tau > 0$ denotes
the largest threshold such that the level set $L(\tau)$ has a probability greater than or equal to $1-\tau$ with
respect to the distribution induced by $f$. Figure \ref{conjuntosnivel} shows the level sets for three different values of the parameter $\tau$. For values of $\tau $ close to one, the level set represents the domain concentrated around the greatest mode. However, if $\tau$ is close to zero, then it represents the effective support of the density $f$.

\begin{figure}[h!] \label{conjuntosnivel}\centering
\hspace{-.3567 cm}\includegraphics[height=4.8cm,
width=0.342\textwidth]{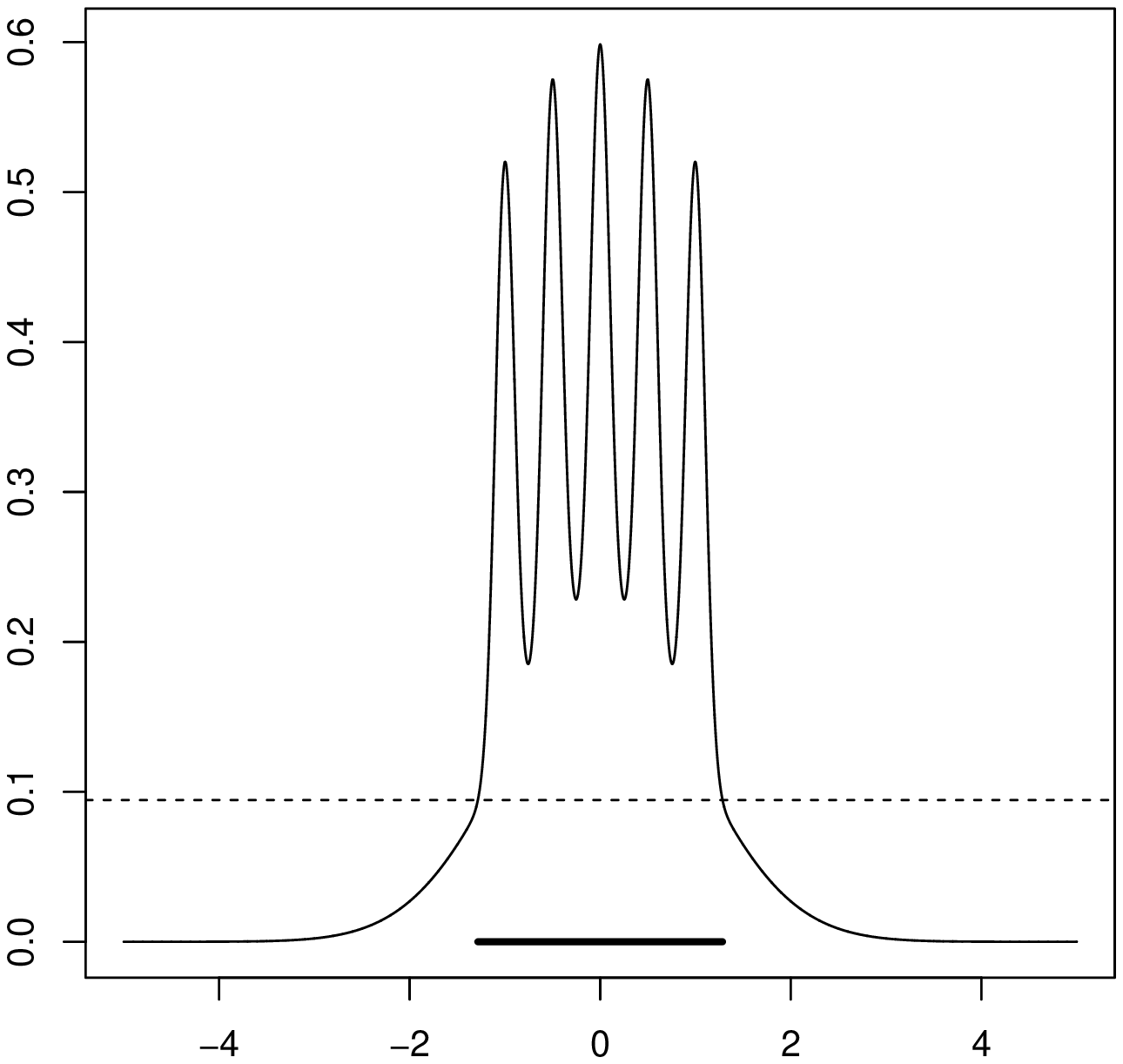}\includegraphics[height=4.8cm, width=0.342\textwidth]{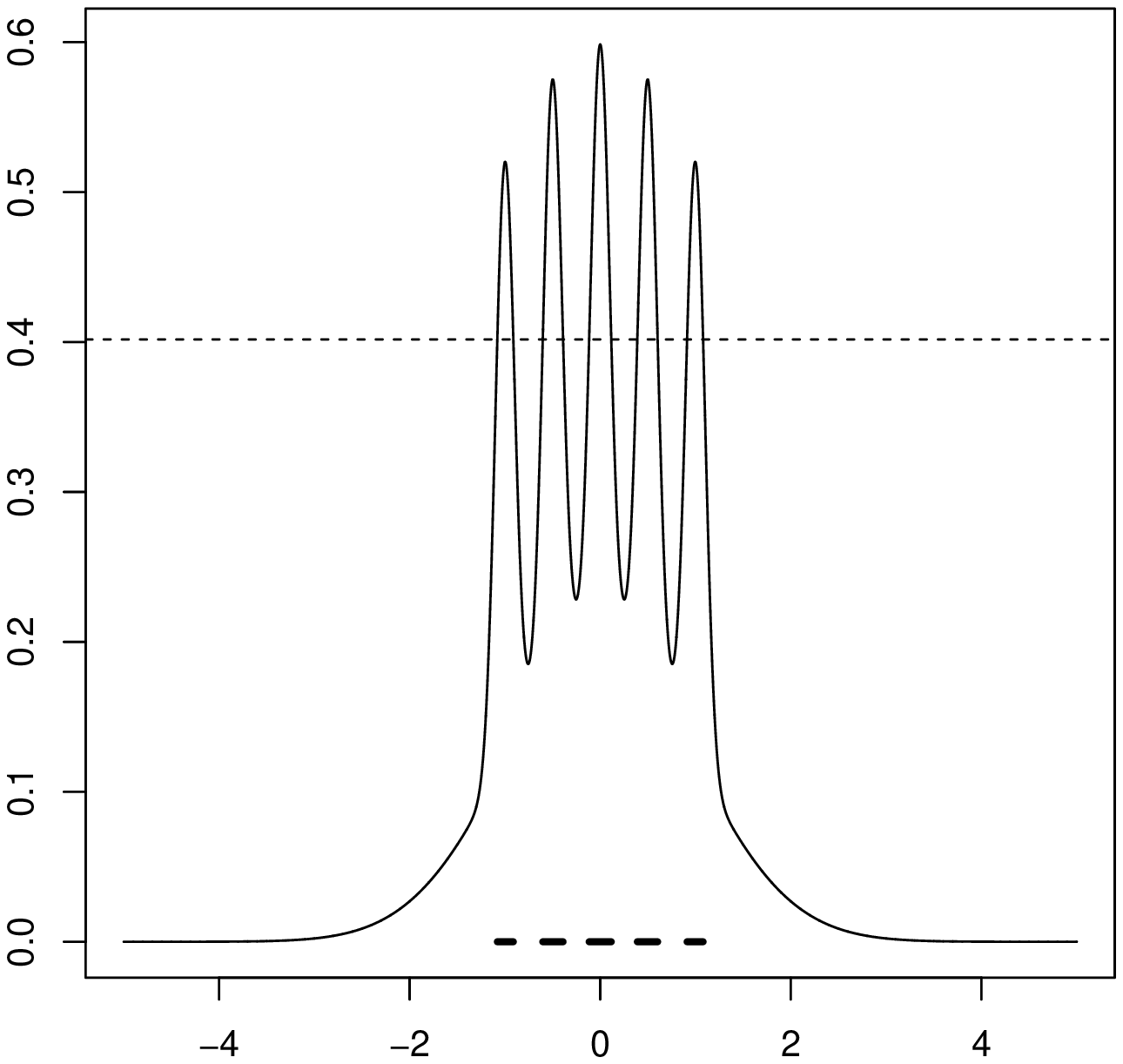}\includegraphics[height=4.8cm, width=0.342\textwidth]{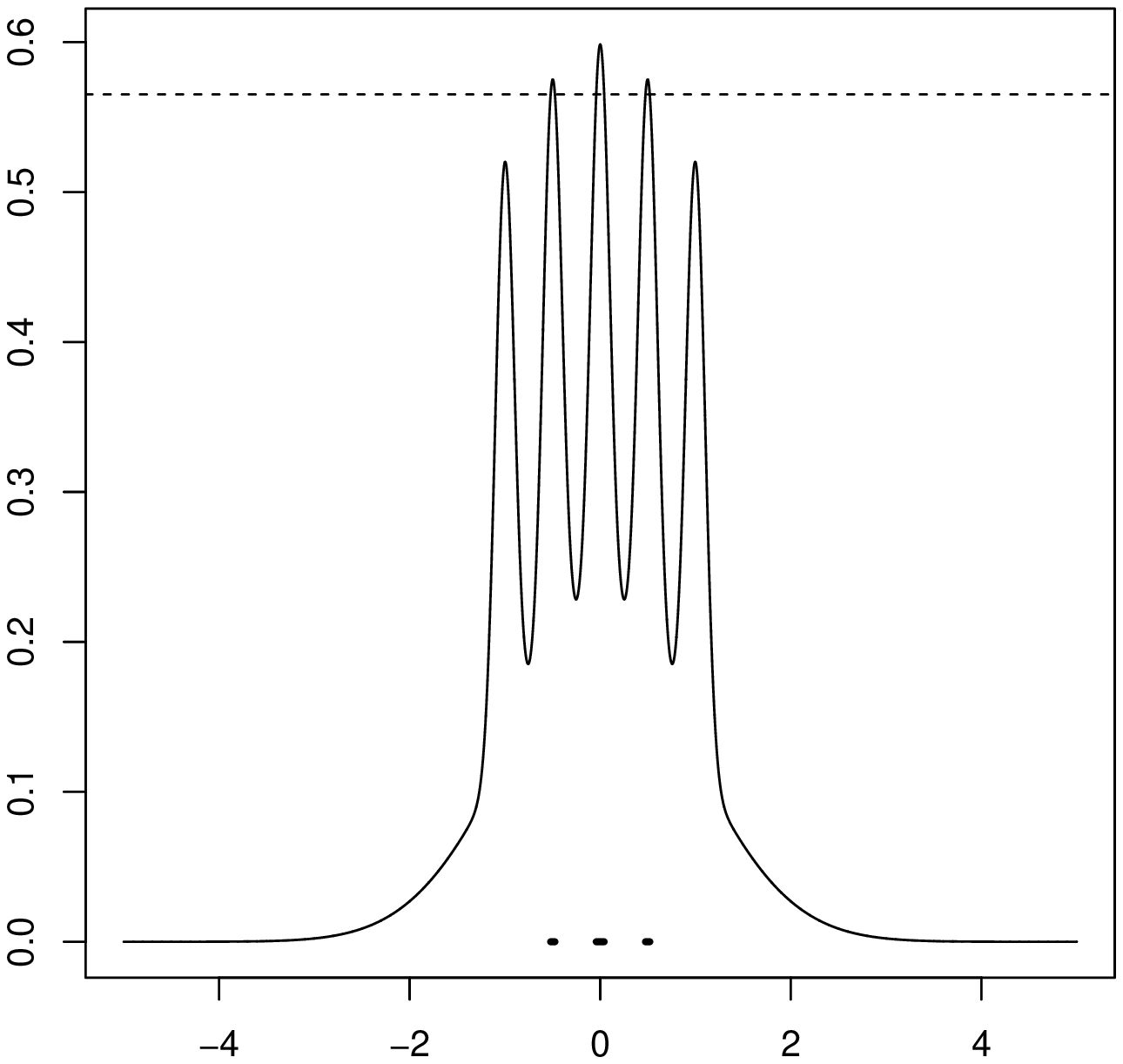}\vspace{-.25cm} \\
\caption{Level sets for a $1-$dimensional density with $\tau=0.1$ (left), $\tau=0.5$ (centre), and $\tau=0.9$ (right).}
\end{figure}\vspace{.35 cm}

Level set estimation plays a crucial role in various scientific fields, and has received considerable interest in the literature. Since Hartigan (1975) introduced the notion of population clusters as the connected
components of density level sets, many interesting applications have appeared. The concept of clustering is related to the notion of the mode and,
in fact, some clustering algorithms are based on the estimation of modes (see Cuevas, Febrero and Fraiman, 2000). An interesting application of this clustering approach to astronomical sky surveys was proposed by Jang (2006), and Klemel\"a (2004, 2006) used
a similar point of view to develop methods for visualizing multivariate density
estimates. Goldenshluger and Zeevi (2004) used level set estimation in the context
of the Hough transform, which is a well-known computer vision algorithm. Some
problems in flow cytometry involve the statistical problem of reconstructing a level
set for the difference between two probability densities (see Roederer and Hardy, 2001). In addition, interesting applications include the detection of mine fields based on aerial observations, the analysis of seismic data, as well as certain issues
in image segmentation (see Huo and Lu, 2004). The detection of outliers is another important application of level set estimation (see Gardner et al. (2006) or Markou and Singh (2003) for a review). An outlier can be thought of as an observation that is in the set $\{f<f_\tau\}$. In other words, the outlier does not belong to the effective support determined by the level set $L(\tau)$. This approach follows that of Devroye and Wise (1980) to determine whether a manufacturing process is out of control. For quality control schemes, see also Ba\'illo, Cuevas and Justel (2000) or Ba\'illo and Cuevas (2006).

%For a Bayesian approach, see Gayraud and Rousseau (2005). In another case,

The broad scope of level set estimation clearly motivates the need
for an in-depth study of the practical performance of existing methods. In general, the problem has been approached using three different nonparametric methodologies: plug-in methods, excess mass methods, and hybrid methods (for a Bayesian alternative, see Gayraud and Rousseau, 2005). This paper focuses on analyzing the behaviour of existing data-driven methods for reconstructing density level sets. The three types of data-driven algorithms for the estimation of level sets will be presented and compared through a detailed simulation study. We focus on the one-dimensional setting so as to include some methods that do not have  multidimensional counterparts, such as the plug-in algorithm proposed by Samworth and Wand (2010) or the excess mass method developed by M\"uller and Sawitzki (1991).

This paper is organized as follows. Plug-in methods are presented and compared in Section \ref{plugin}. Sections \ref{exceso de masa} and \ref{hybridos} discuss the behaviour of excess mass and hybrid methods, respectively. In addition, two new hybrid algorithms are proposed in Section 4. The most competitive methods in each group are then  compared in Section \ref{comparacionfinal}. Finally, we present some general conclusions in Section \ref{conc}.

}
\section{Comparative study of plug-in methods}\label{plugin}

\normalsize{
The simplest option for estimating level sets is the so-called plug-in methodology. This is based on replacing the unknown density $f$ by a suitable nonparametric estimator $f_n$, usually that of the kernel. Given $\mathcal{X}_n$, the kernel density estimator at a point $x$ is defined as\vspace{.1 cm}
$$f_n(x)=\frac{1}{nh}\sum_{i=1}^n K\left(\frac{X_i-x}{h}\right),\vspace{.1 cm}$$
where $K$  denotes a kernel function (in what follows, the Gaussian kernel) and $h>0$ denotes the bandwidth parameter. Thus, this group of methods proposes $\hat{L}(\tau)=\{f_n\geq\hat{f}_\tau\}$ as an estimator of $L(\tau)$, where $\hat{f}_\tau$ estimates $f_\tau$ as follows:
\begin{enumerate}
\item Through numerical integration methods by solving \vspace{.1cm}
$$\int_{\{f_n\geq t \}}f_n(x)dx=1-\tau \vspace{.1cm}$$
for $t$. This algorithm is consistent with the plug-in philosophy, as $\hat{f}_\tau$ is the corresponding threshold if $f$ is replaced by $f_n$. However, this may be inefficient from a computational point of view. For consistency results, see Cadre (2006).
\item Hyndman (1996) proposed a method for estimating $f_\tau$ by calculating the $\tau-$quantile of the empirical distribution of $f_n(X_1),...,f_n(X_n)$. The computational cost of this approach is lower than that of the previous method. For consistency results, see Cadre, Pelletier and Pudlo (2013).
\end{enumerate}

The plug-in methodology is the most common approach, and has received considerable attention in the literature, e.g., Tsybakov (1997), Ba\'illo (2003), Mason and Polonik (2009), Rigollet and Vert (2009), or Mammen and Polonik (2013). However, its performance is heavily dependent on the choice of the bandwidth parameter $h$ for estimating $f$, and there are many
methods of selecting $h$ in the density estimation. Next, we review existing specific bandwidth selectors for density level set estimation.

Ba\'illo and Cuevas (2006) used quality control ideas to estimate level sets. In this context, $L(\tau)$ can be seen as a population tolerance region. Let us assume that a machine working in normal operation produces a sequence
of independent observations $\mathcal{X}_n$ drawn from the density $f$. For example, $X_i$ could be a value for a certain quality characteristic of a manufactured product. If the process starts to run out of control, the distribution of the samples will change. The aim is to detect a real change in this distribution as soon as possible, subject to the bound $\tau \in (0,1)$ on the rate of false alarms. The key idea is that, if there is no change in the distribution for a new observation, then it is most likely to be within the tolerance limits of the region $L(\tau)$. In practice, we may determine that the $n+1$-th observation is a change-point in the distribution of the process, that is, a new observation $X_{n+1}$ does not follow the distribution of $\mathcal{X}_n$ if $X_{n+1}$ does not belong to the plug-in estimator of $L(\tau)$. Hence, choosing a good smoothing parameter to reconstruct a level set in the context of quality control is an interesting problem. Cuevas and Ba\'illo (2006) proposed a bandwidth selector by minimizing a cross-validation estimate of $|\mathbb{P}_{f_n}(h)-\tau| \mbox{ where }
\mathbb{P}_{f_n}(h)=\int_{\{f_n<\hat{f}_\tau\}}f$ denotes the probability of a false alarm. Specifically, $\mathbb{P}_{f_n}(h)$ is approximated by\vspace{.1 cm}
$$ \mathbb{P}_{CV}(h)=\frac{1}{n}\sum_{i=1}^n \mathbb{I}_{\{f_{n,-i}(X_i)<\hat{f}_{\tau,-i}\}},\vspace{.13 cm}$$
where $f_{n,-i}$ is the kernel estimator with bandwidth $h$, constructed from $X_1,...,X_{i-1},X_{i+1},...,X_{n}$, and $\hat{f}_{\tau,-i}$ satisfies\vspace{.1 cm}
$$\int_{\{f_{n,-i}\geq\hat{f}_{\tau,-i}\}}f_{n,-i}=1-\tau.\vspace{.13 cm}$$

Samworth and Wand (2010) proposed a new automatic rule to select the smoothing parameter in the one-dimensional case. Their proposal is based on a uniform-in-bandwidth asymptotic approximation of a specific set estimation risk function, $\mathbb{E}\{d_{\mu_f}(L(\tau),\hat{L}(\tau))\}$, where\vspace{.1 cm} $$d_{\mu_f}(L(\tau),\hat{L}(\tau))=\int_{L(\tau)\triangle \hat{L}(\tau)}f(t)\mbox{ }dt\vspace{.1 cm}$$
and $\triangle$ denotes the symmetric difference given by $L(\tau)\triangle \hat{L}(\tau)=(L(\tau) \setminus \hat{L}(\tau))\cup(\hat{L}(\tau) \setminus L(\tau))$. This asymptotic approximation of the risk function was used to develop a plug-in bandwidth
selection strategy by minimizing an estimator of the asymptotic risk. It was proved that, under some smoothness assumptions on $f$, the solution to this minimization problem is, with probability one, unique for large $n$. The asymptotic risk depends on the threshold $f_\tau$, the density function, and its first and second derivatives ($f^{'}$ and $f^{''}$, respectively) evaluated at the unknown extremes $x_1,...,x_{2r}$ of the intervals that form the theoretical level set. These three functions are estimated using kernel estimators with different pilot bandwidths $\hat{h}_0$, $\hat{h}_1$, and $\hat{h}_2$ for $f$, $f^{'}$, and $f^{''}$, respectively. The kernel estimators can then be denoted by $f_{n,\hat{h}_0}$, $f_{n,\hat{h}_1}^{'}$, and $f_{n,\hat{h}_2}^{''}$. The threshold $f_\tau$ is estimated by numerical integration from $f_{n,\hat{h}_0}$, whereas the points $x_1,...,x_{2r}$ are estimated by solving the equation $f_{n,\hat{h}_0}(x)=\hat{f}_\tau$. However, it should be noted that this algorithm can provide level set estimators that are equal to the empty set, mainly for large values of $\tau$. The reason for this is simple---the estimation of the pilot bandwidth $\hat{h}_0$, which is used to estimate the threshold $f_\tau$, can be considerably greater than the final smoothing parameter $\hat{h}_{SW}$ obtained by Samworth and Wand's algorithm. The density estimator calculated from $\hat{h}_{SW}$ does not then intersect $\hat{f}_\tau$. To solve this problem, the threshold must be recalculated from $\hat{h}_{SW}$ after obtaining it. The modified version of this method will be considered in the rest of this paper.

Singh, Scott and Nowak (2009) presented a plug-in procedure for reconstructing density level sets based on a histogram. This method considers a collection of cells $A$ as a regular partition of $[0, 1]^d$ ($d\geq 1$) into hypercubes $\mathcal{A}_j$ of dyadic sidelength $2^{-j}$, where $j$ is a nonnegative integer. The estimator of the level set $C(\lambda)=\{f\geq \lambda\}$ for a given value of $\lambda>0$ at a resolution level of $j$ is defined as\vspace{.12 cm}
$$\hat{C}(\lambda)=\bigcup_{\{A\in \mathcal{A}_j:f_{n,H}(A)\geq \lambda\}}A \mbox{ with }f_{n,H}(A)=\frac{\mathbb{P}_n(A)}{\mu(A)},\vspace{.12 cm}$$
where $ \mathbb{P}_n$ denotes the empirical probability induced by $\mathcal{X}_n$ and $\mu$ is the Lebesgue measure. $L(\tau)$ can be estimated by $\hat{C}(\hat{f}_\tau)$ where, to avoid the problem of bandwidth selection, $\hat{f}_\tau$ is computed using the empirical procedure proposed by Walther (1997). That is,\vspace{.12 cm}
$$\hat{f}_\tau =\max\{ \lambda>0: \mathbb{P}_n(\hat{C}(\lambda))\geq 1-\tau \}.\vspace{.12 cm}$$
The algorithm suggested in Singh, Scott and Nowak (2009) depends on the resolution level $j$. This parameter is selected using a data-driven procedure (see Section 3.2 in Singh, Scott and Nowak, 2009). It was proved that near-minimax optimal rates of convergence for a specific class of level sets are achieved for the resulting density level set estimator.

Of course, it is also possible to consider classical methods such as cross-validation (see Bowman, 1984) or plug-in algorithms (see Sheather and Jones, 1991) to select the bandwidth parameter, although these are not specific to the estimation of level sets.
}

%Simulation results for these methods are shown in Section \ref{pluginsimu}.

\subsection{Simulations results for plug-in methods}\label{pluginsimu}

\normalsize{
In this section, we compare the methods of Ba\'illo and Cuevas (BC), Samworth and Wand (SW), Sheather and Jones (SJ), and cross-validation (CV). BC and SW are specific bandwidth selectors for estimating level sets. The latter two algorithms are general selectors for the estimation of density functions. The behaviour of these general methods has been studied in the context of density level set estimation, e.g., in Cao, Cuevas and Gonz\'alez-Manteiga (1994).  The results of Singh, Scott and Nowak's algorithm are not included in this comparison because, for the sample size considered in the study, its behaviour was not satisfactory.

%The next notation will be used to refer to the different methods: Samworth and Wand's method will be denoted by SW, Ba\'illo and Cuevas' method by BC, cross validation selector by CV and Sheather and Jones selector by SJ.

$ $

{

\centering
\includegraphics[height=2.9cm, width=0.21\textwidth]{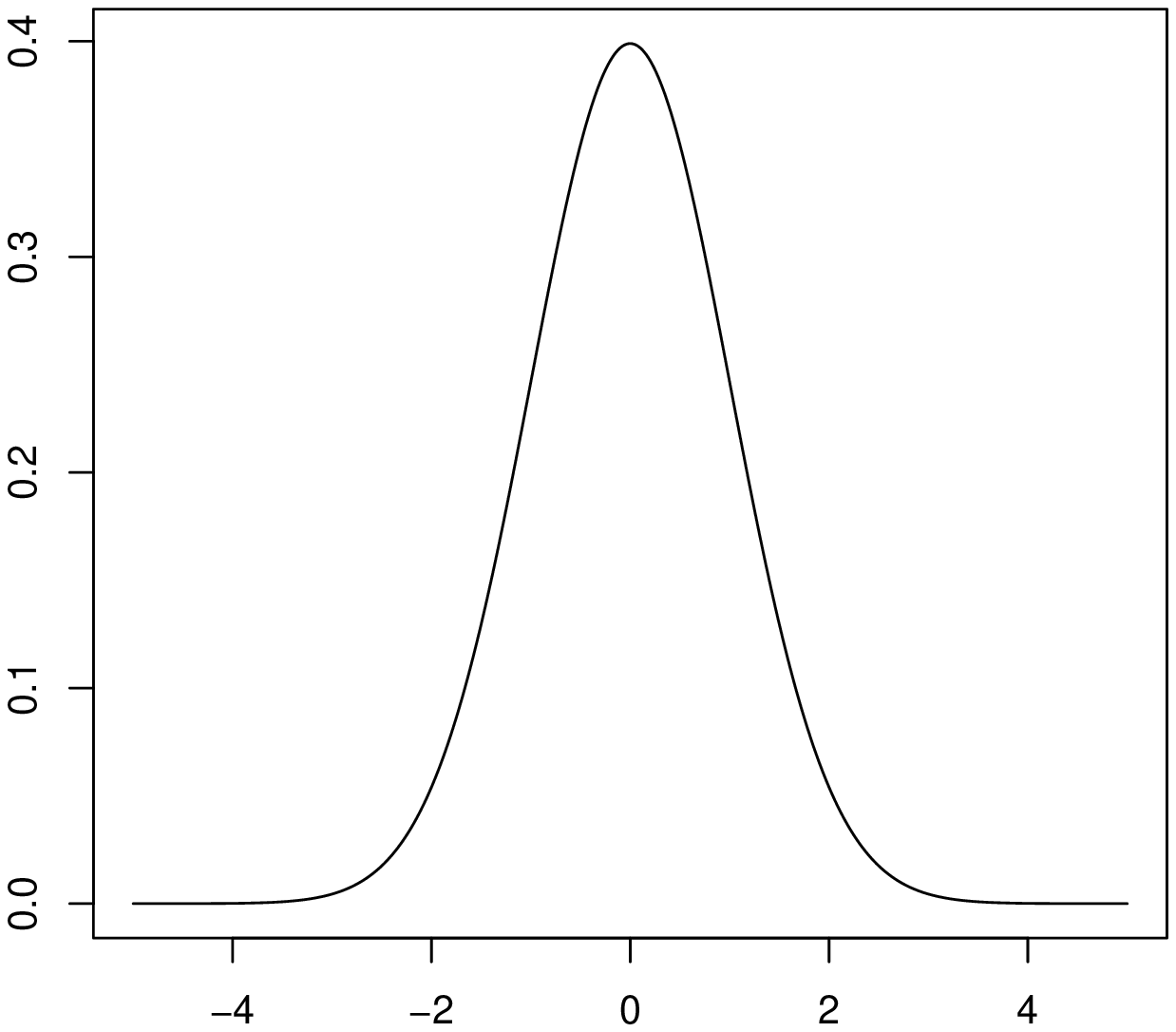} \includegraphics[height=2.9cm, width=0.21\textwidth]{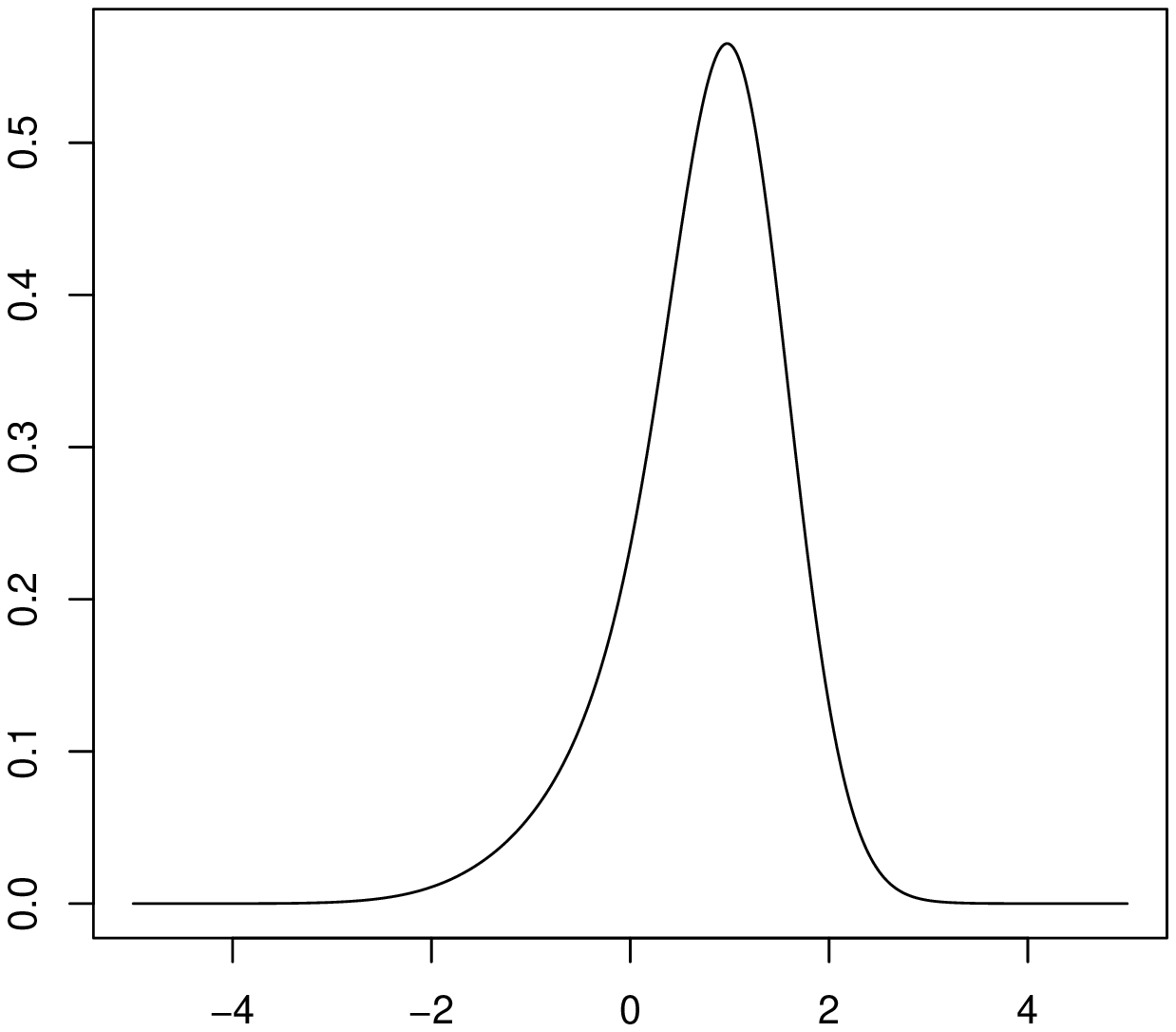} \includegraphics[height=2.9cm, width=0.21\textwidth]{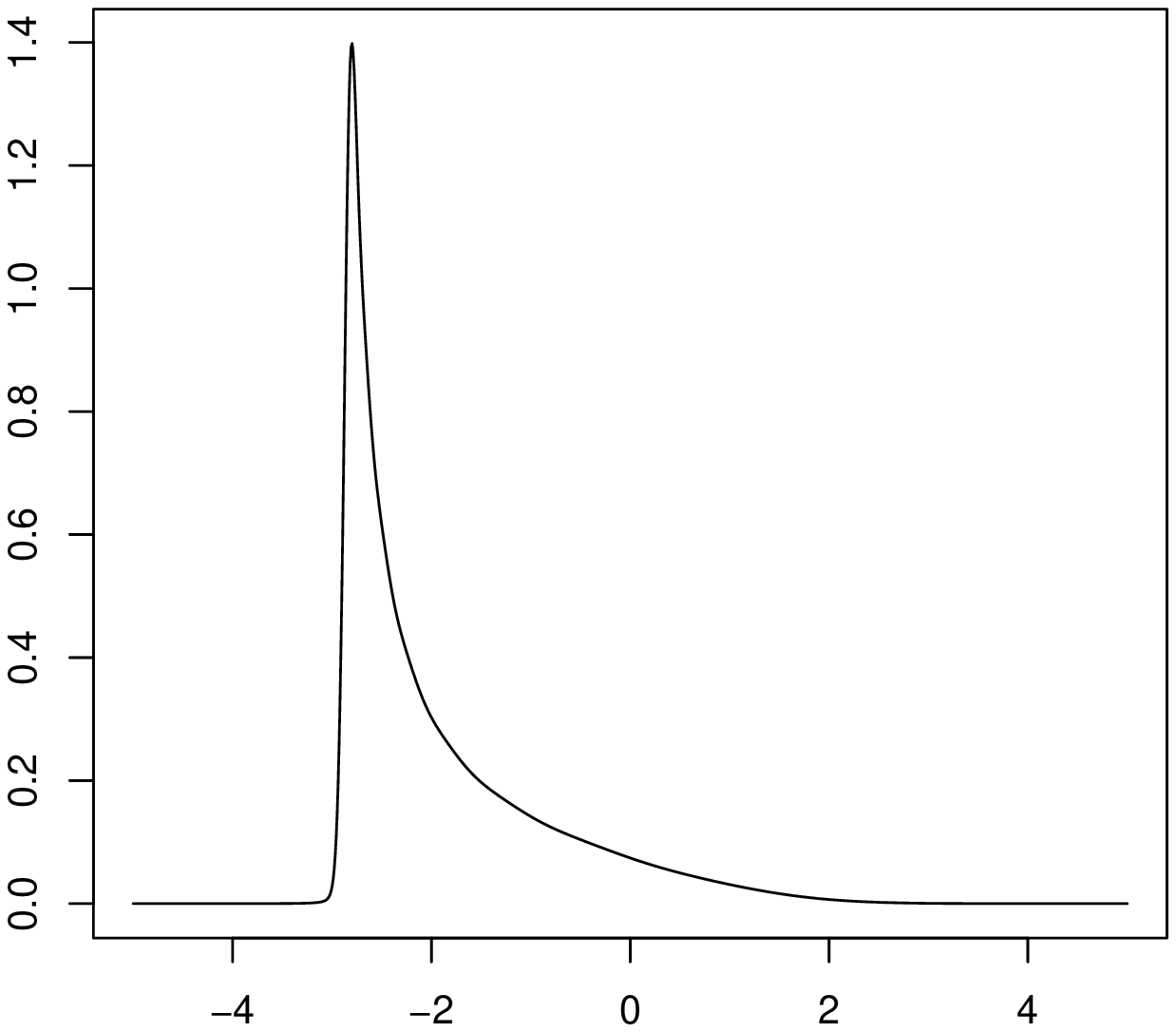}\vspace{-.5cm}\\
\centering
\includegraphics[height=2.9cm, width=0.21\textwidth]{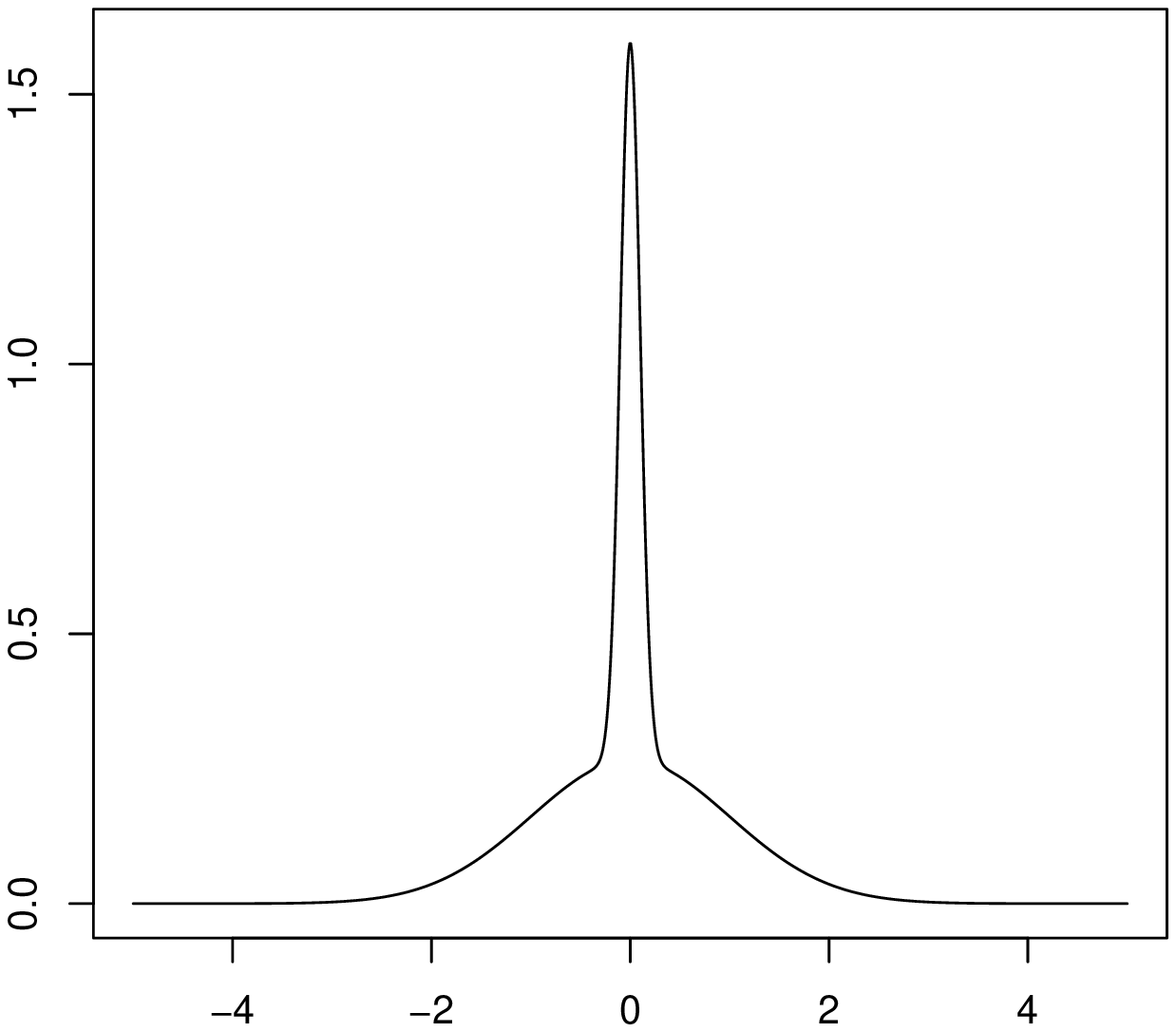} \includegraphics[height=2.9cm, width=0.21\textwidth]{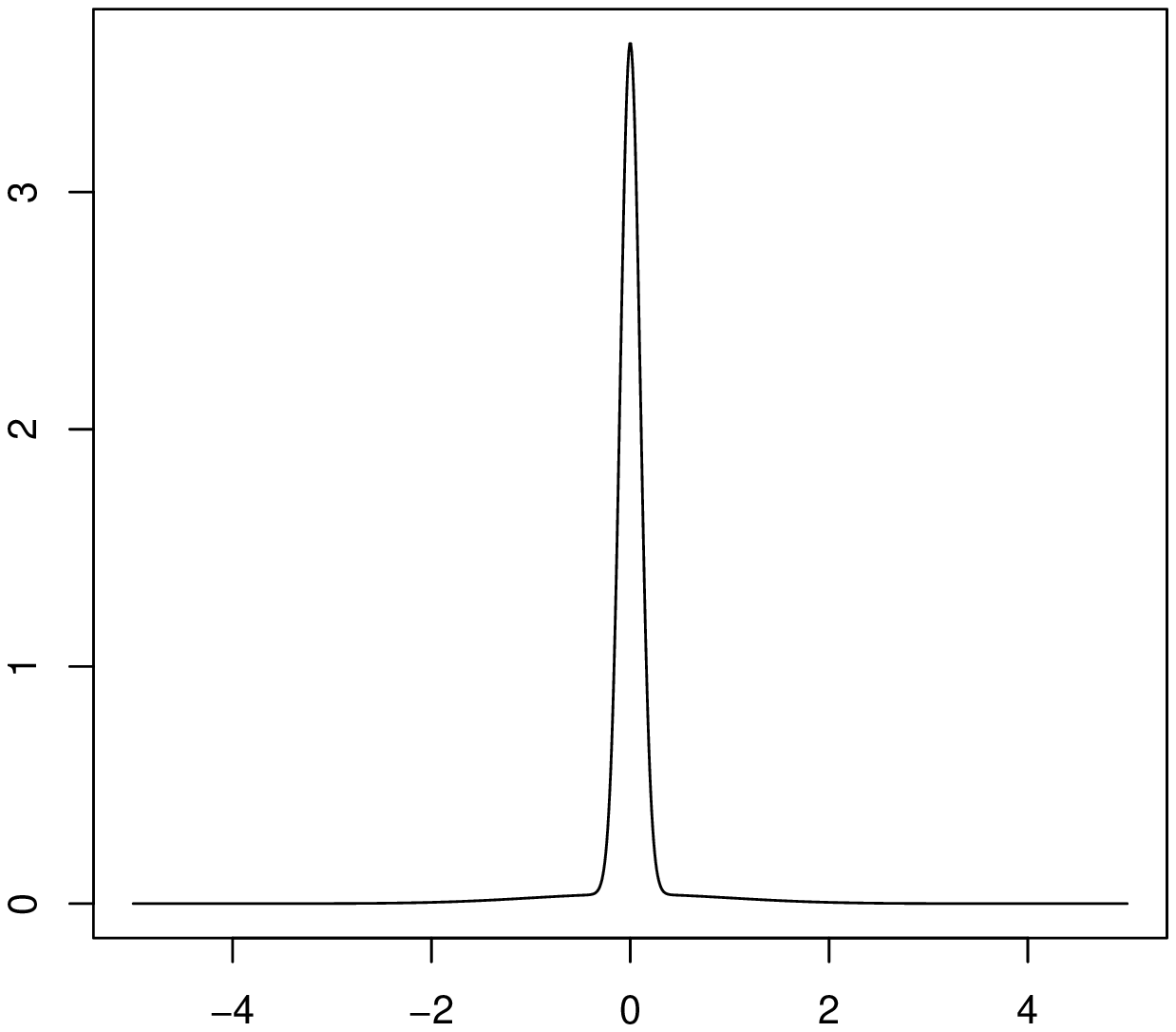} \includegraphics[height=2.9cm, width=0.21\textwidth]{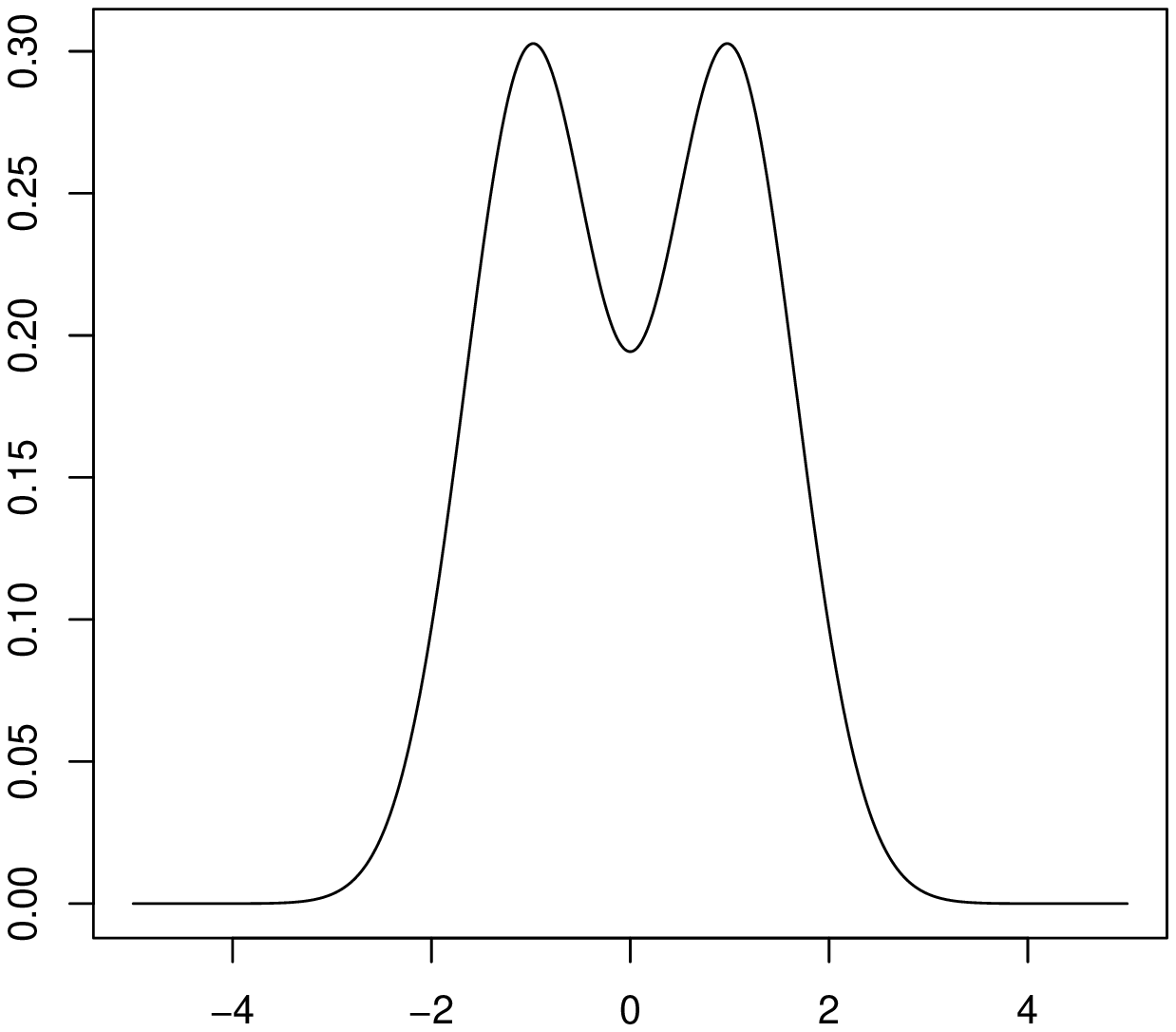}\vspace{-.5cm}\\

\centering
\includegraphics[height=2.9cm, width=0.21\textwidth]{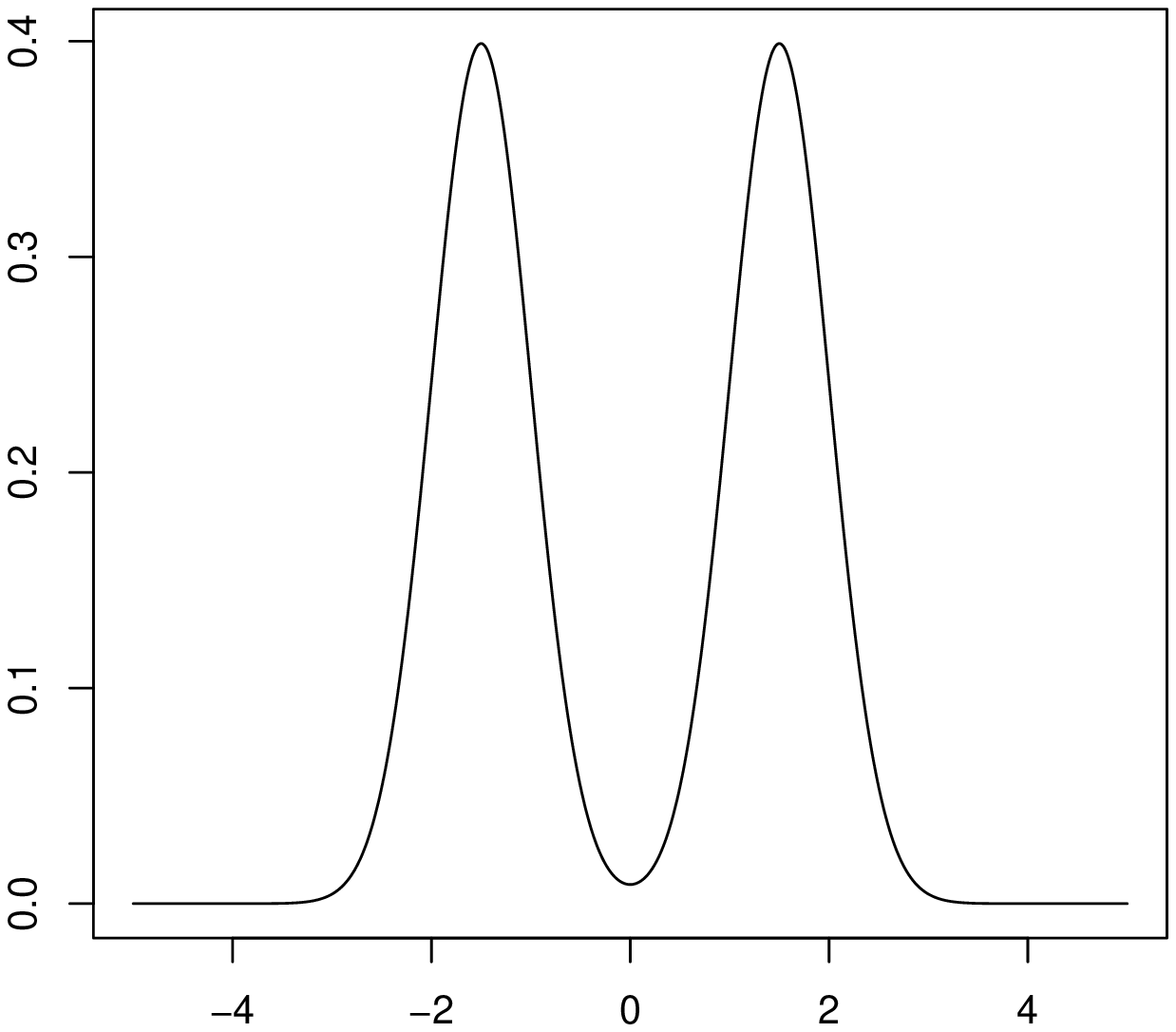} \includegraphics[height=2.9cm, width=0.21\textwidth]{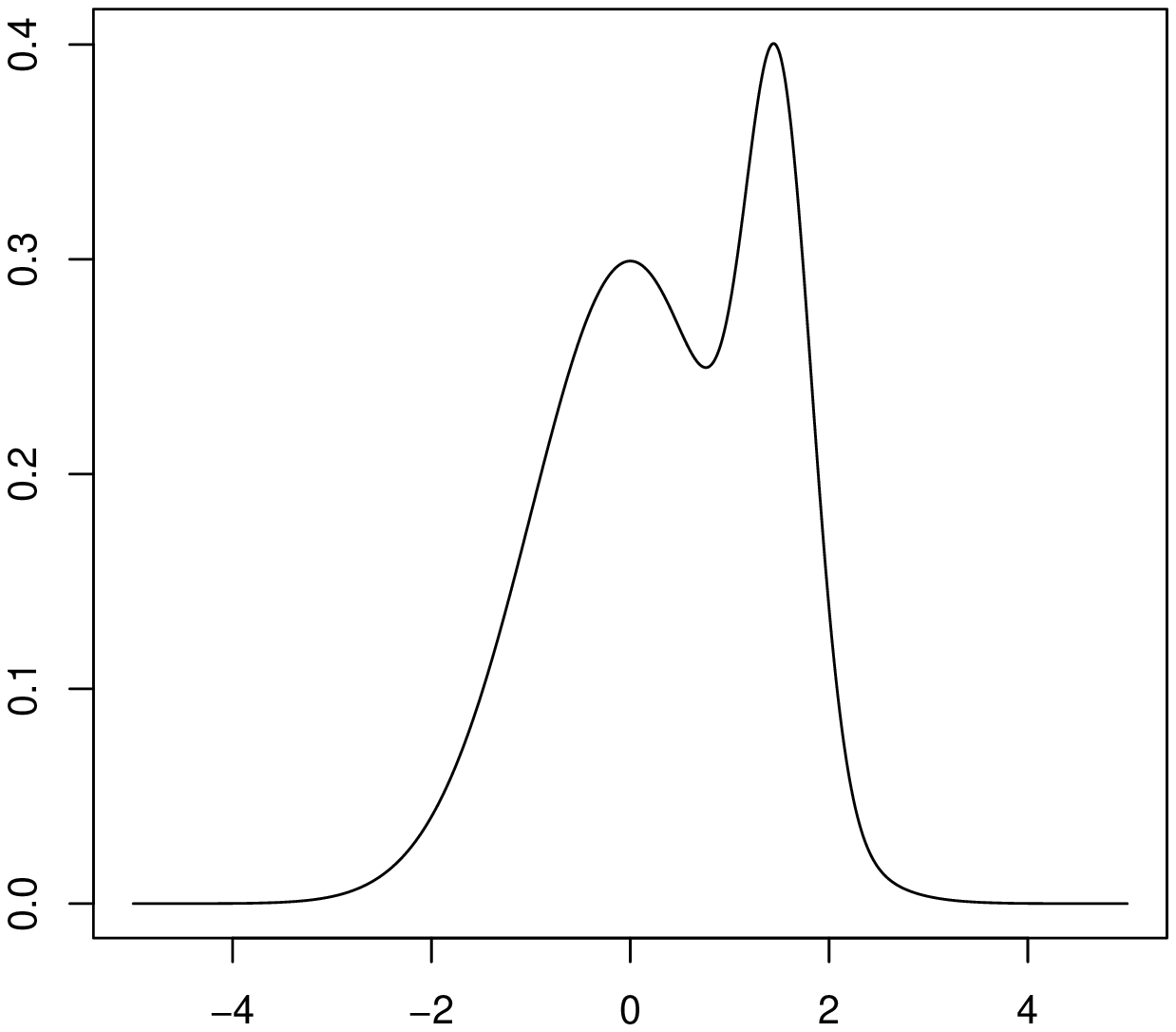} \includegraphics[height=2.9cm, width=0.21\textwidth]{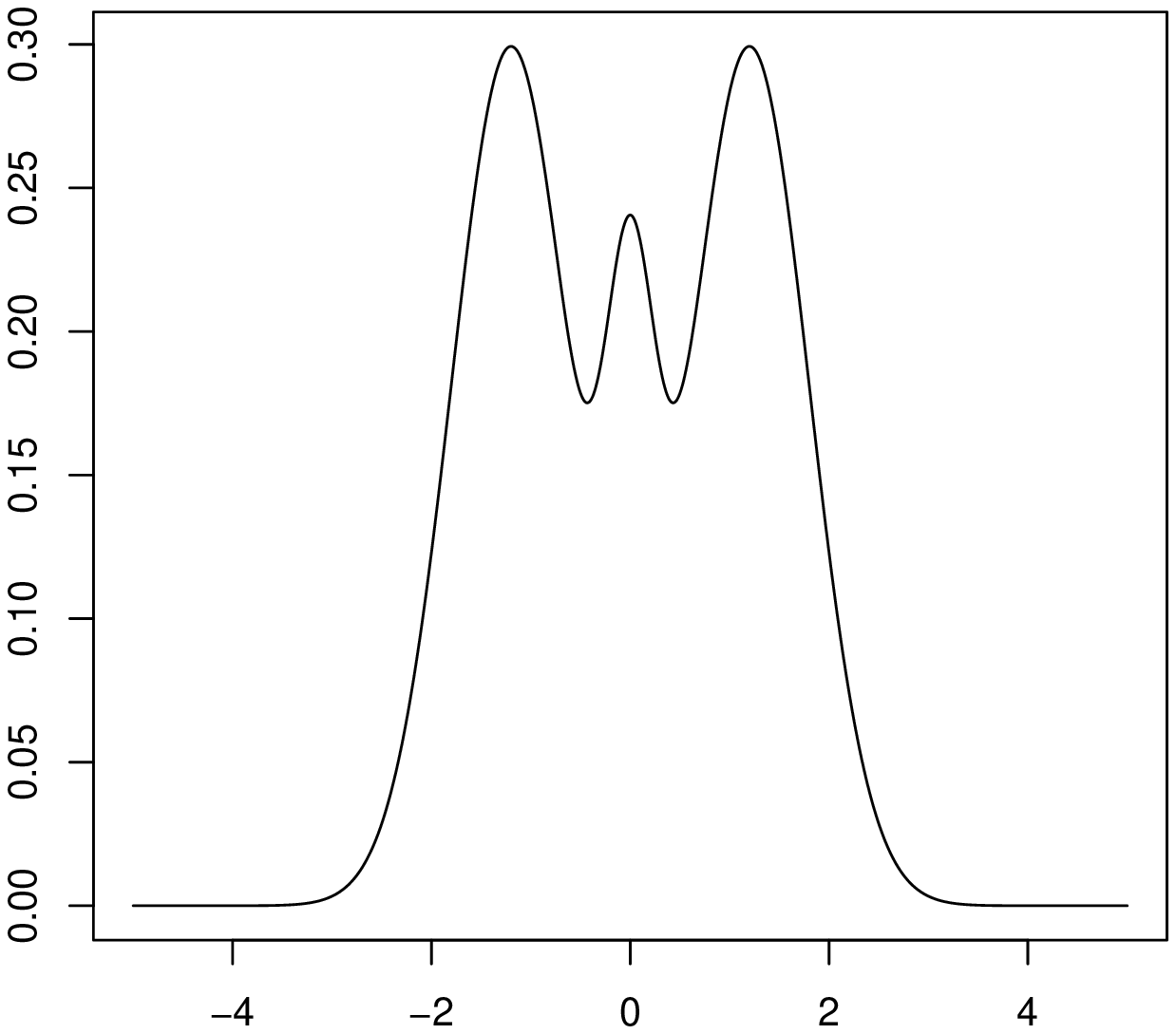}\vspace{-.5cm}\\

\centering
\includegraphics[height=2.9cm, width=0.21\textwidth]{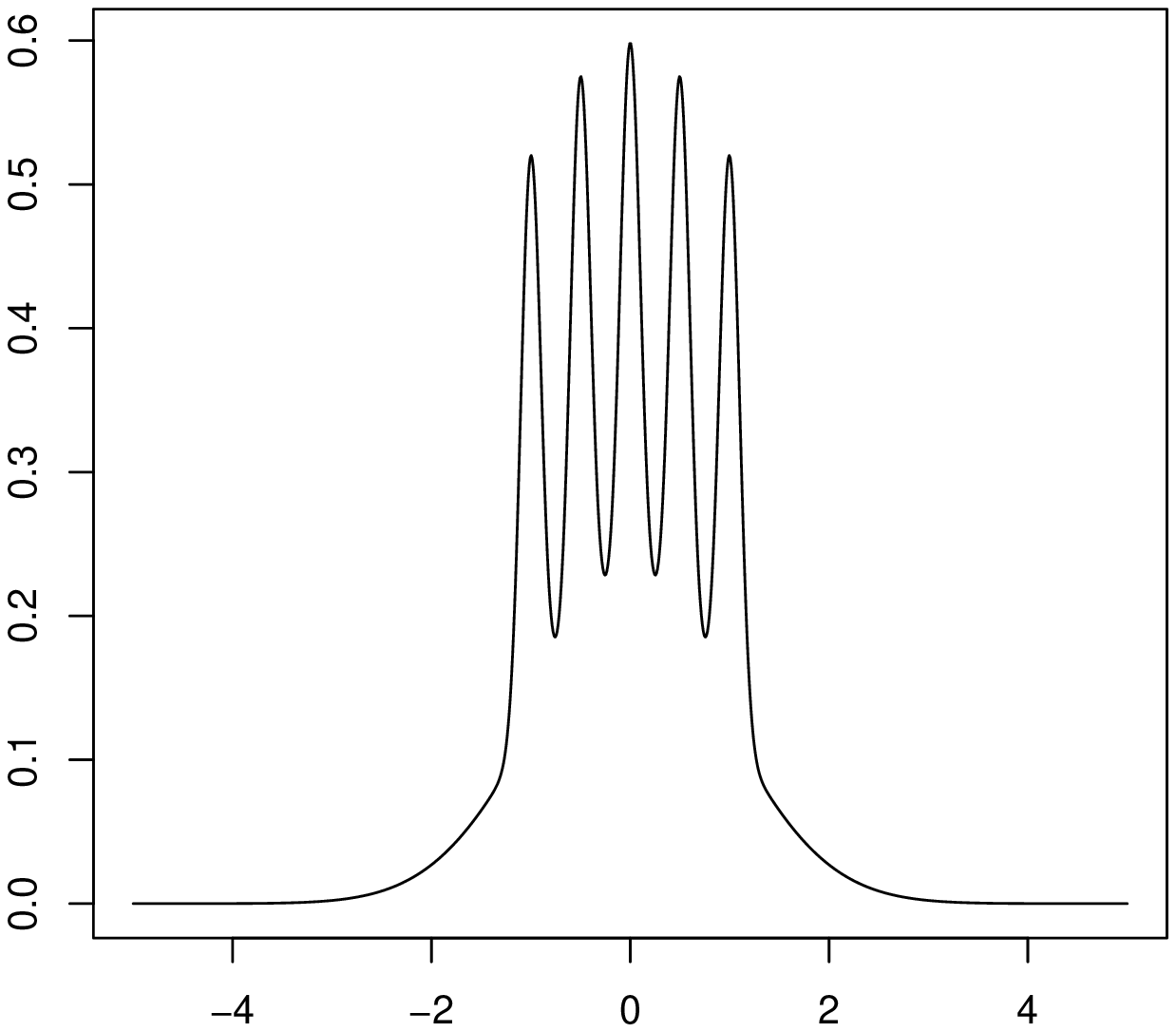} \includegraphics[height=2.9cm, width=0.21\textwidth]{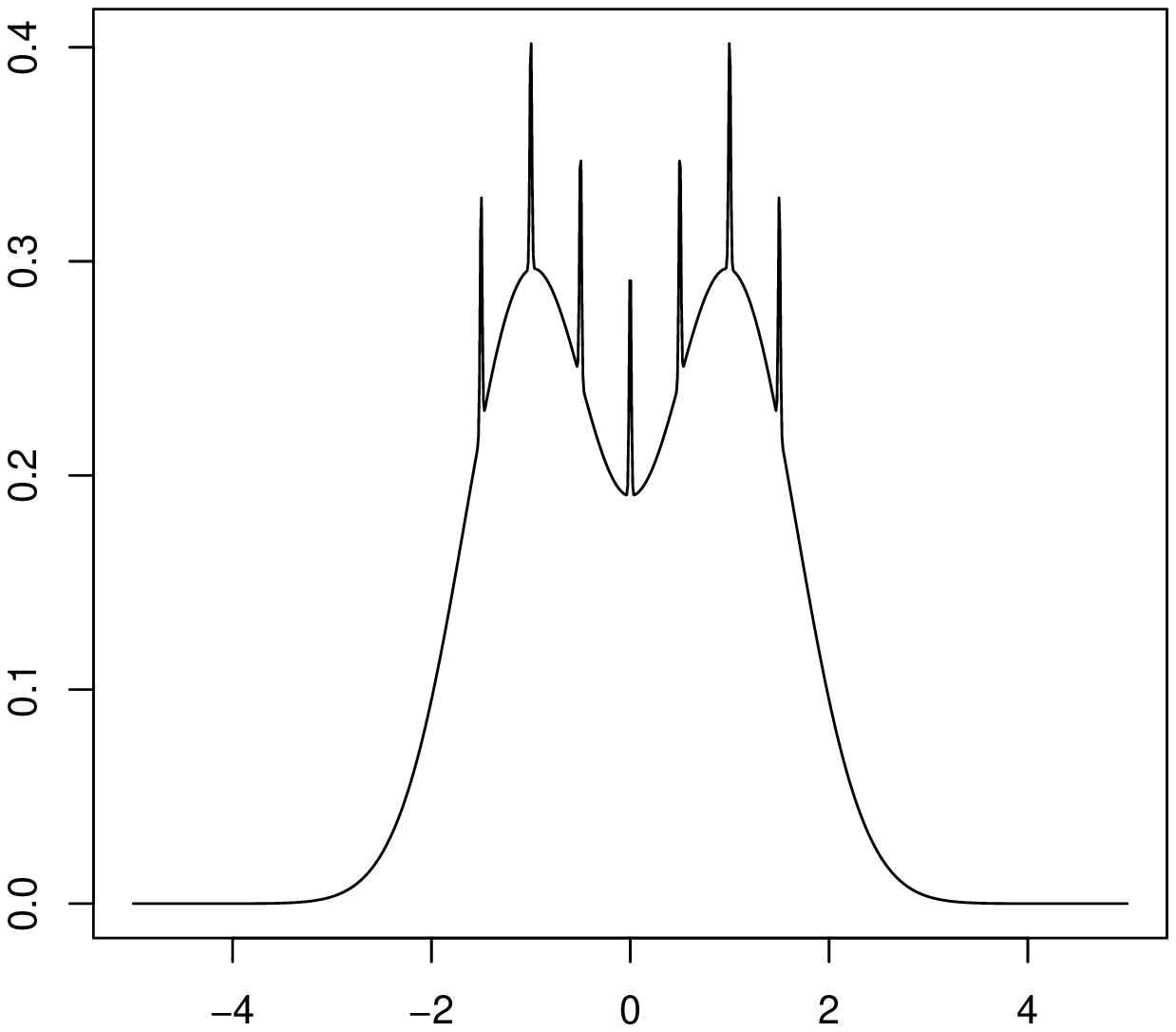} \includegraphics[height=2.9cm, width=0.21\textwidth]{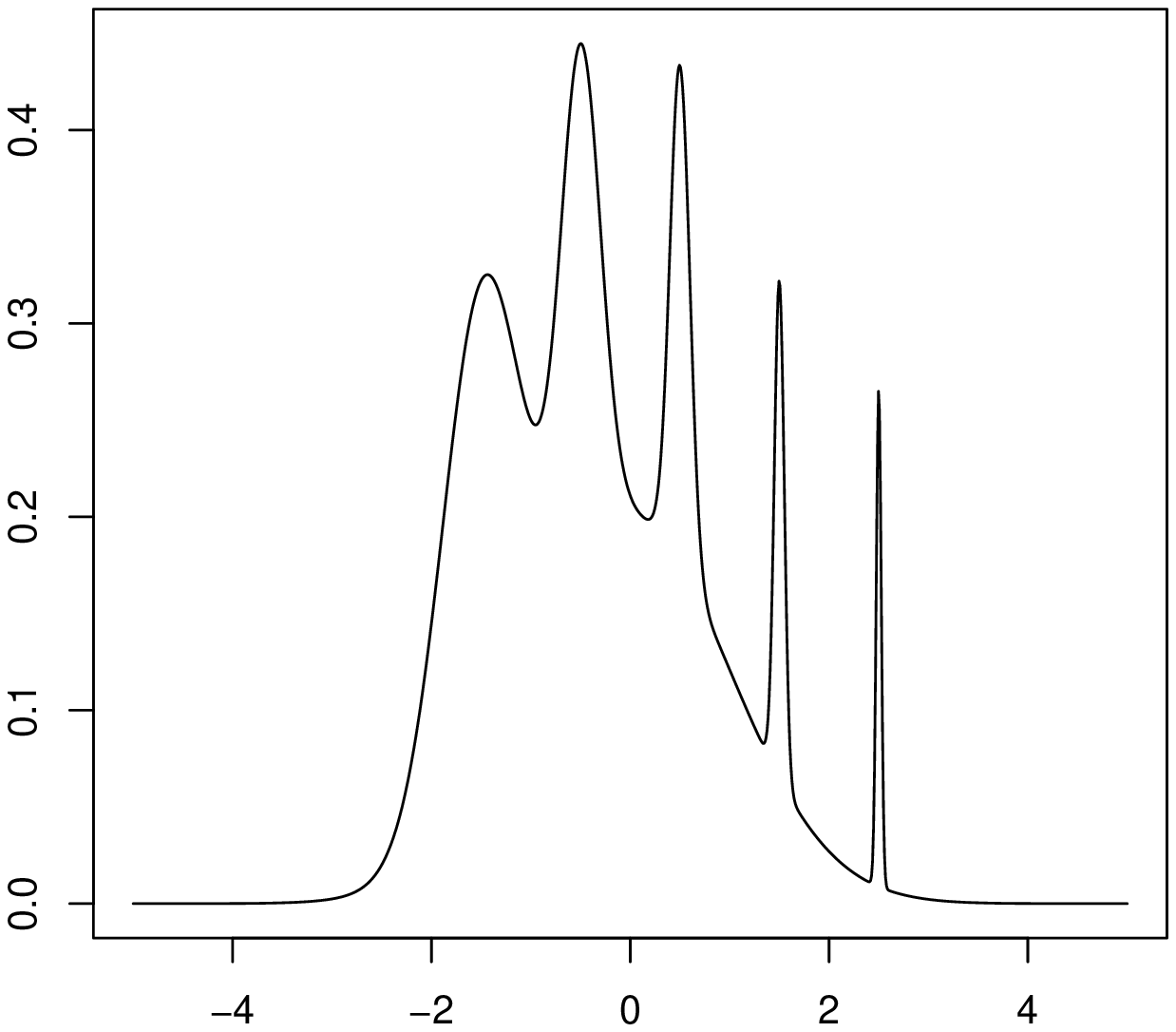}\vspace{-.5cm}\\

\centering
\includegraphics[height=2.9cm, width=0.21\textwidth]{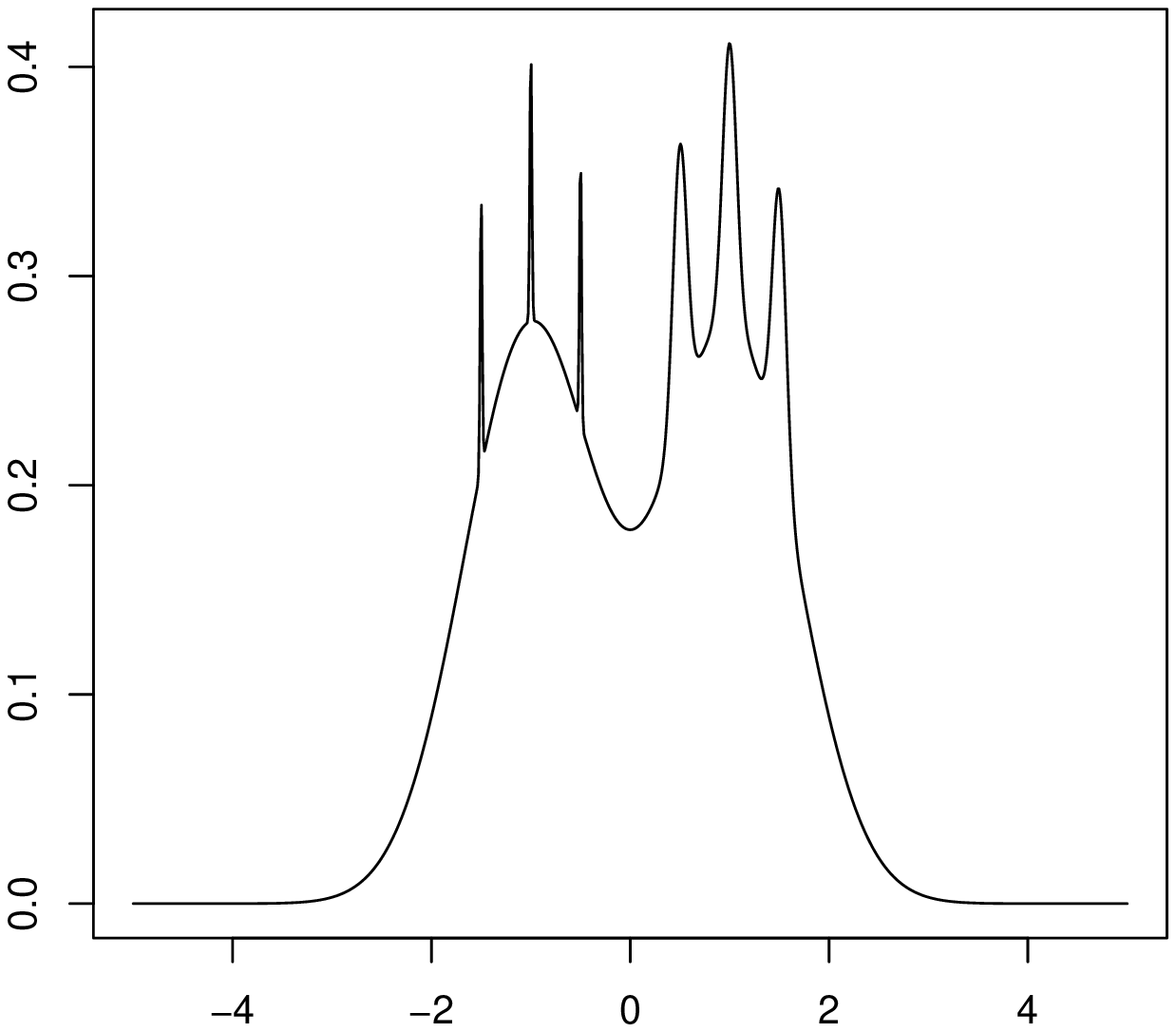} \includegraphics[height=2.9cm, width=0.21\textwidth]{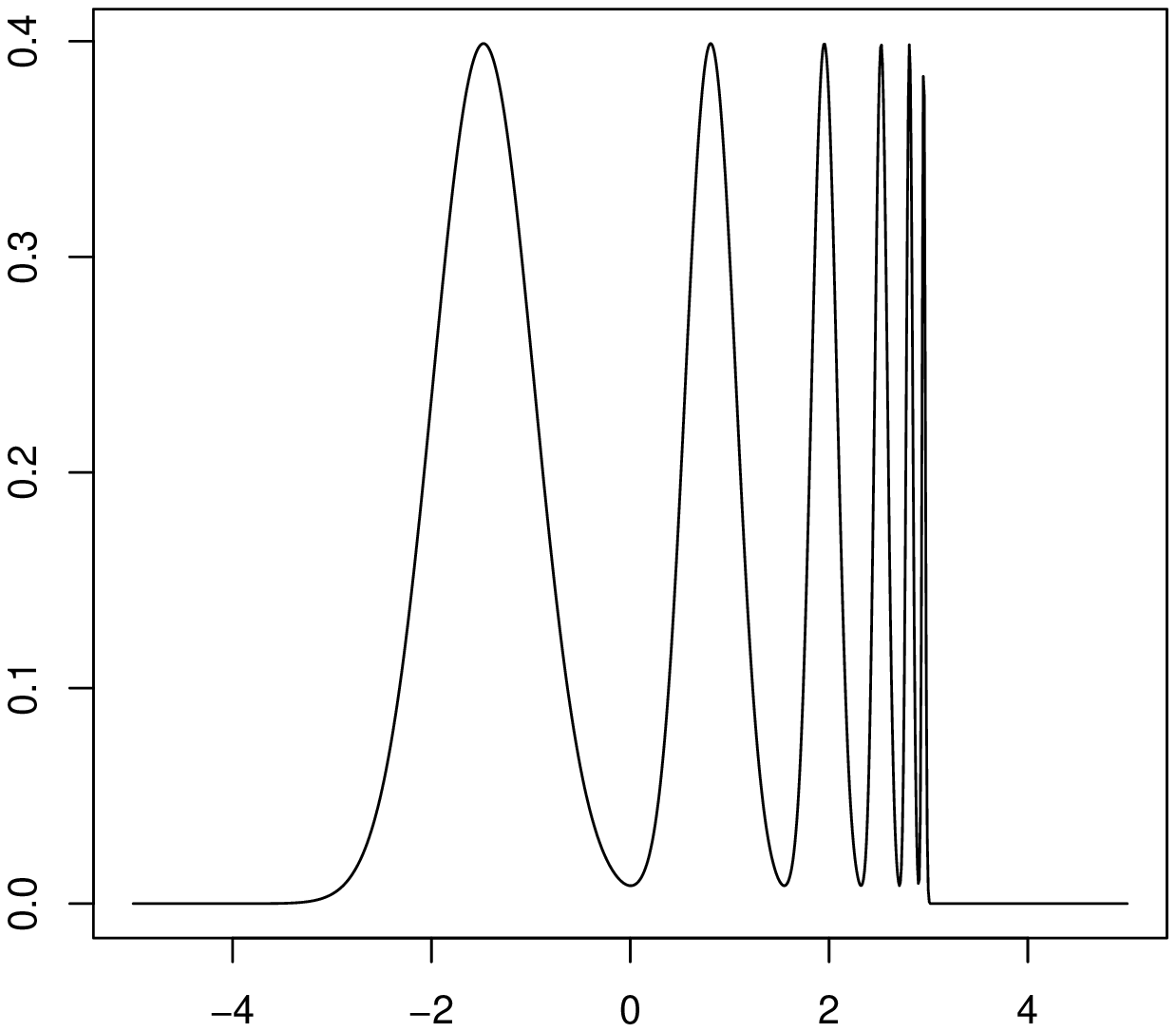} \includegraphics[height=2.9cm, width=0.21\textwidth]{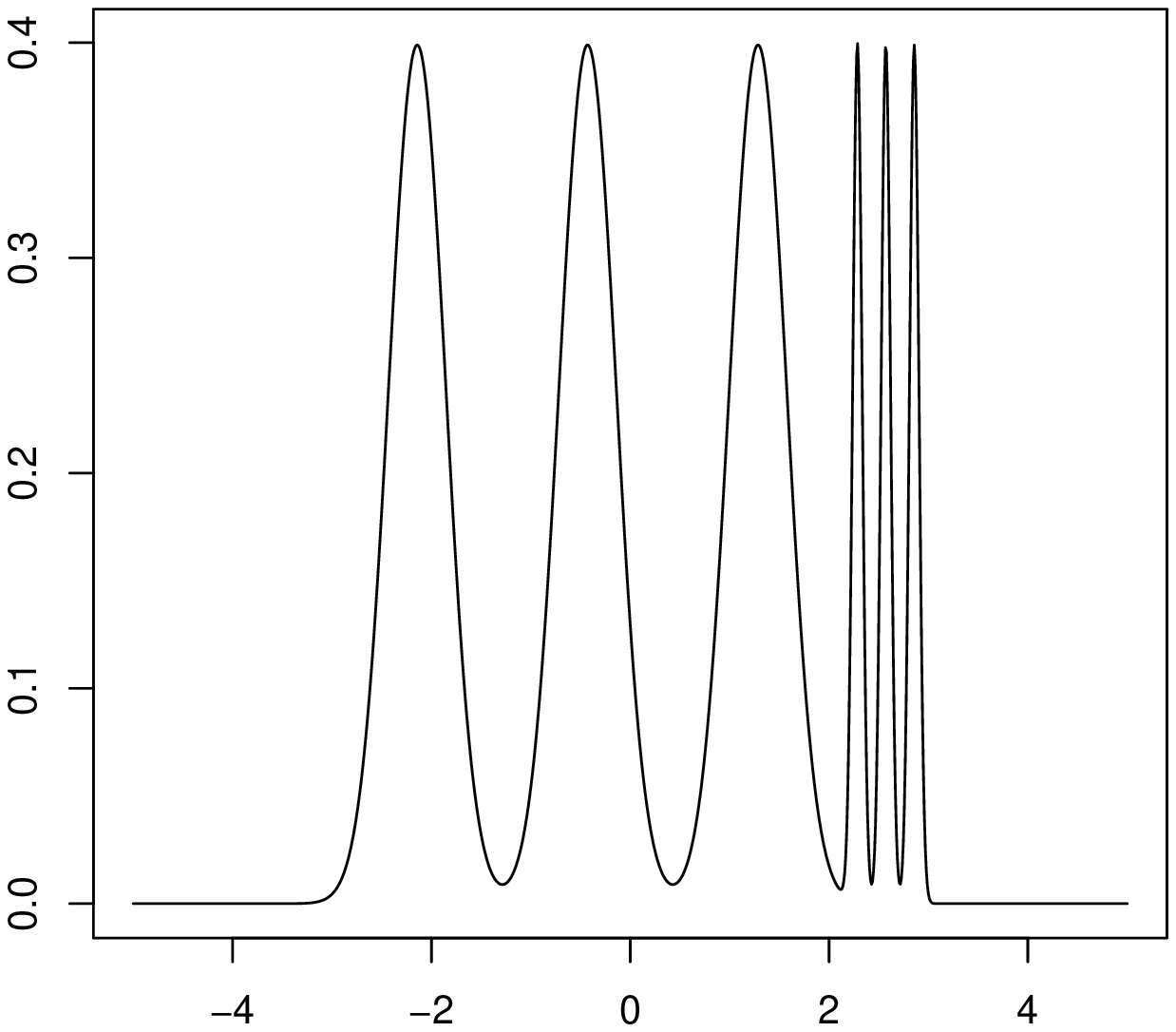}\vspace{-.5cm}\\

\centering
\includegraphics[height=2.9cm, width=0.21\textwidth]{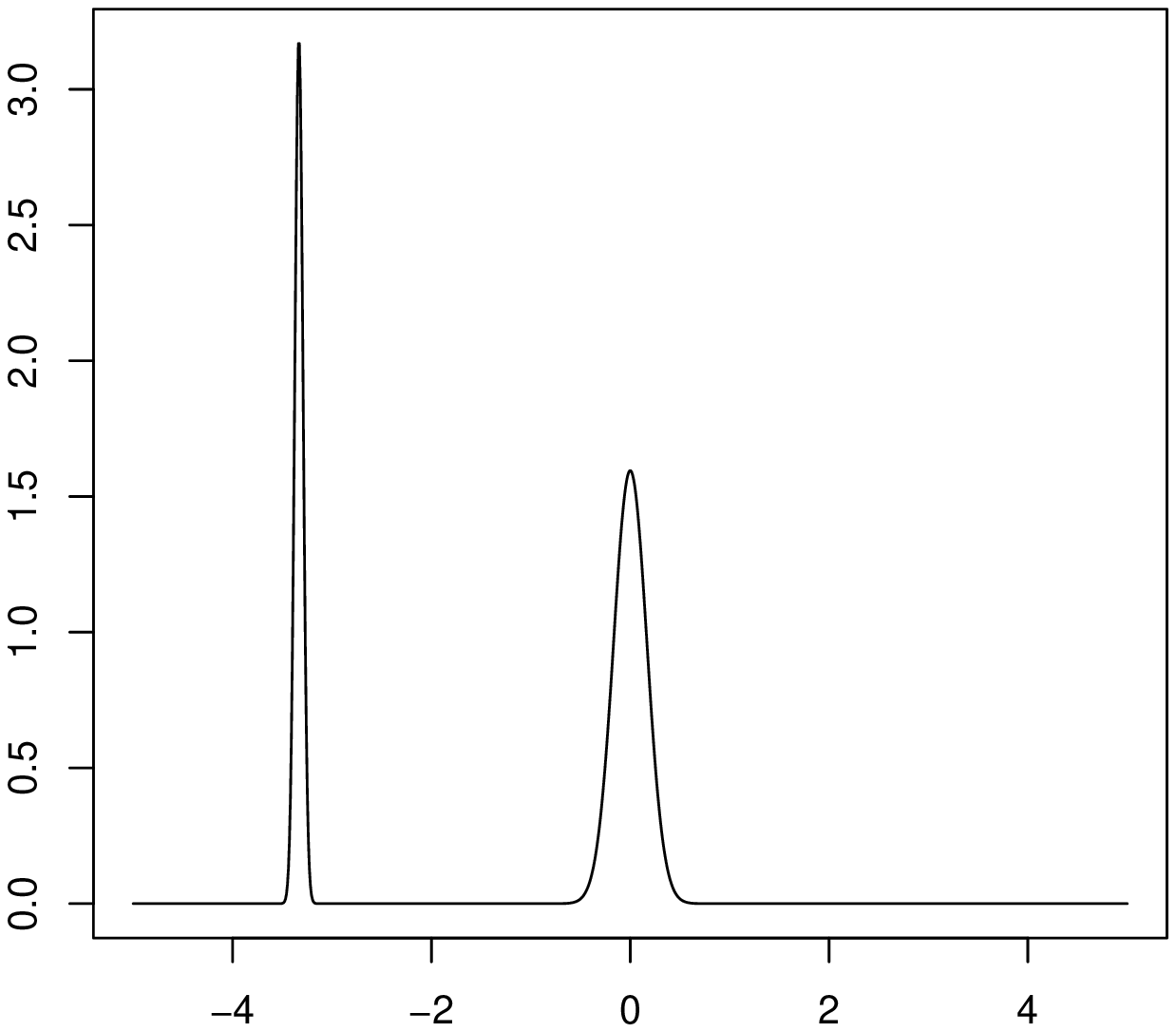} \includegraphics[height=2.9cm, width=0.21\textwidth]{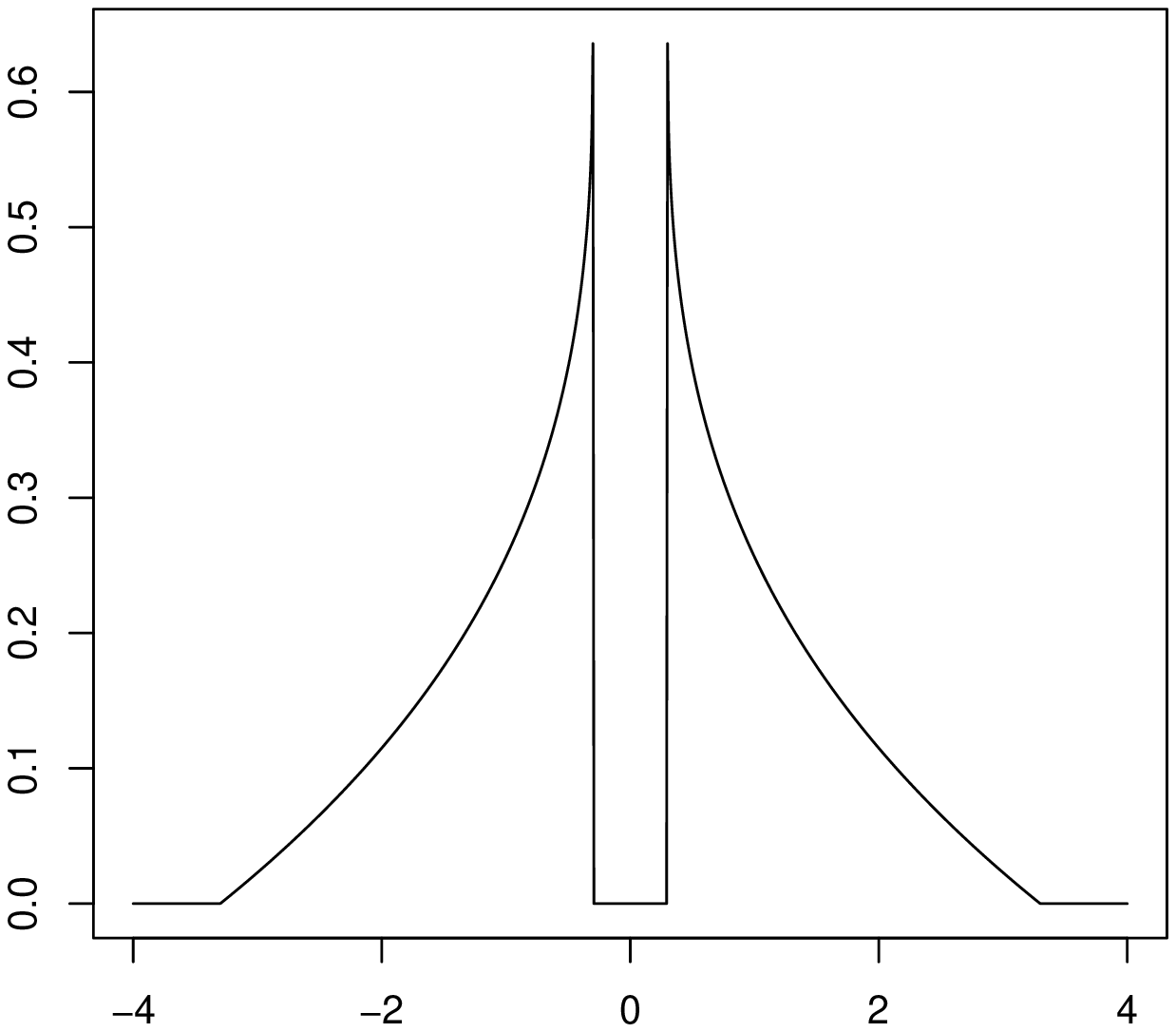} \includegraphics[height=2.9cm, width=0.21\textwidth]{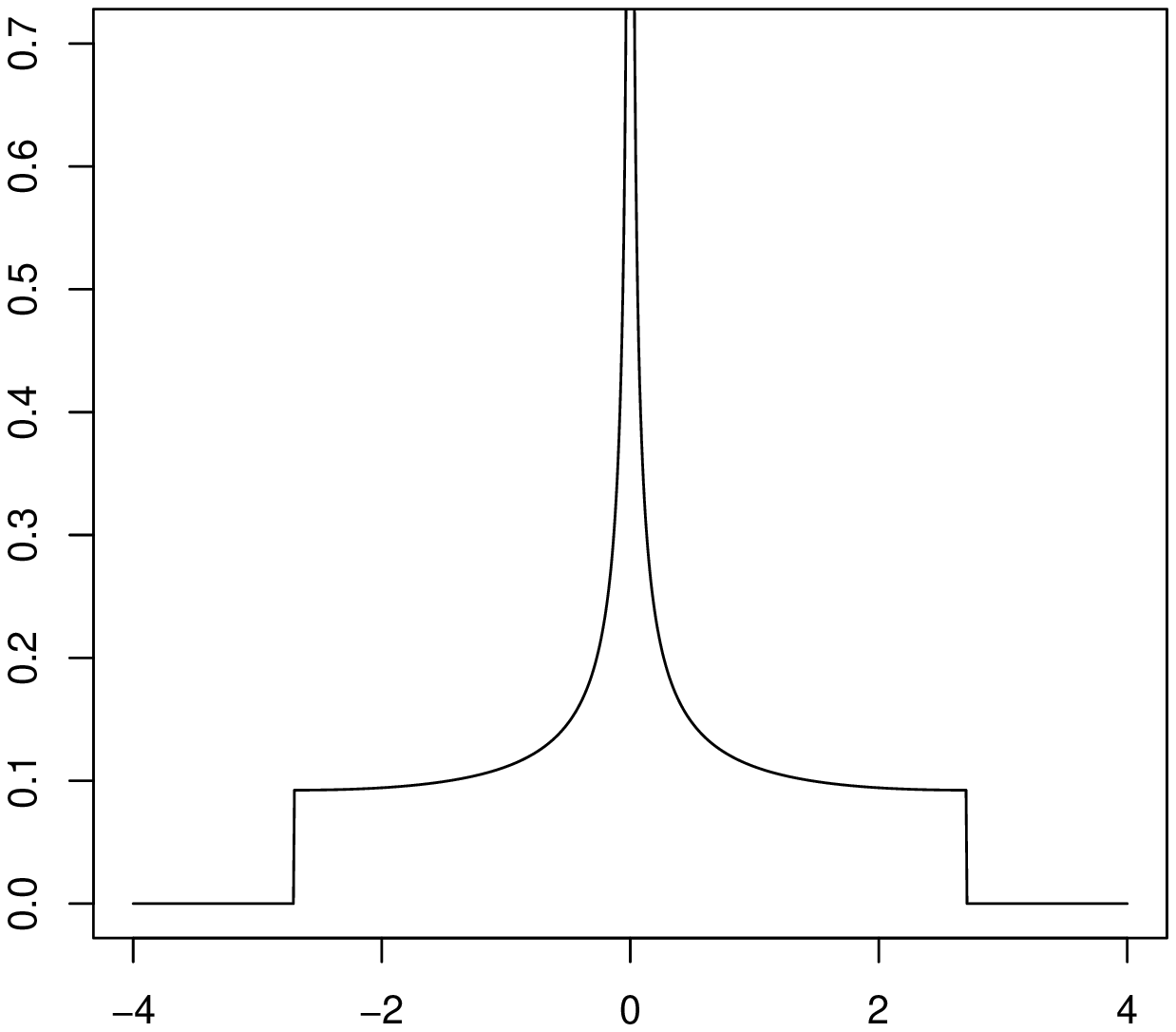}\vspace{-.9 cm}\\

}

\begin{figure}[h]
\centering
\begin{picture}(0,0)
\put(-57,363){\begin{sideways}\tiny{\textbf{Model 1}}\end{sideways}}
\put(-57,296){\begin{sideways}\tiny{\textbf{Model 4}}\end{sideways}}
\put(-57,225){\begin{sideways}\tiny{\textbf{Model 7}}\end{sideways}}
\put(-57,152){\begin{sideways}\tiny{\textbf{Model 10}}\end{sideways}}
\put(-57,83){\begin{sideways}\tiny{\textbf{Model 13}}\end{sideways}}
\put(-57,13){\begin{sideways}\tiny{\textbf{Model 16}}\end{sideways}}
\put(28.8,363){\begin{sideways}\tiny{\textbf{Model 2}}\end{sideways}}
\put(28.8,296){\begin{sideways}\tiny{\textbf{Model 5}}\end{sideways}}
\put(28.8,225){\begin{sideways}\tiny{\textbf{Model 8}}\end{sideways}}
\put(28.8,152){\begin{sideways}\tiny{\textbf{Model 11}}\end{sideways}}
\put(28.8,83){\begin{sideways}\tiny{\textbf{Model 14}}\end{sideways}}
\put(28.8,13){\begin{sideways}\tiny{\textbf{Model 17}}\end{sideways}}
\put(114.9,363){\begin{sideways}\tiny{\textbf{Model 3}}\end{sideways}}
\put(114.9,296){\begin{sideways}\tiny{\textbf{Model 6}}\end{sideways}}
\put(114.9,225){\begin{sideways}\tiny{\textbf{Model 9}}\end{sideways}}
\put(114.9,152){\begin{sideways}\tiny{\textbf{Model 12}}\end{sideways}}
\put(114.9,83){\begin{sideways}\tiny{\textbf{Model 15}}\end{sideways}}
\put(114.9,13){\begin{sideways}\tiny{\textbf{Model 18}}\end{sideways}}
\end{picture}\vspace{.3cm}
\normalsize{\caption{Density models for the simulation study.}\label{iok}}
\end{figure}
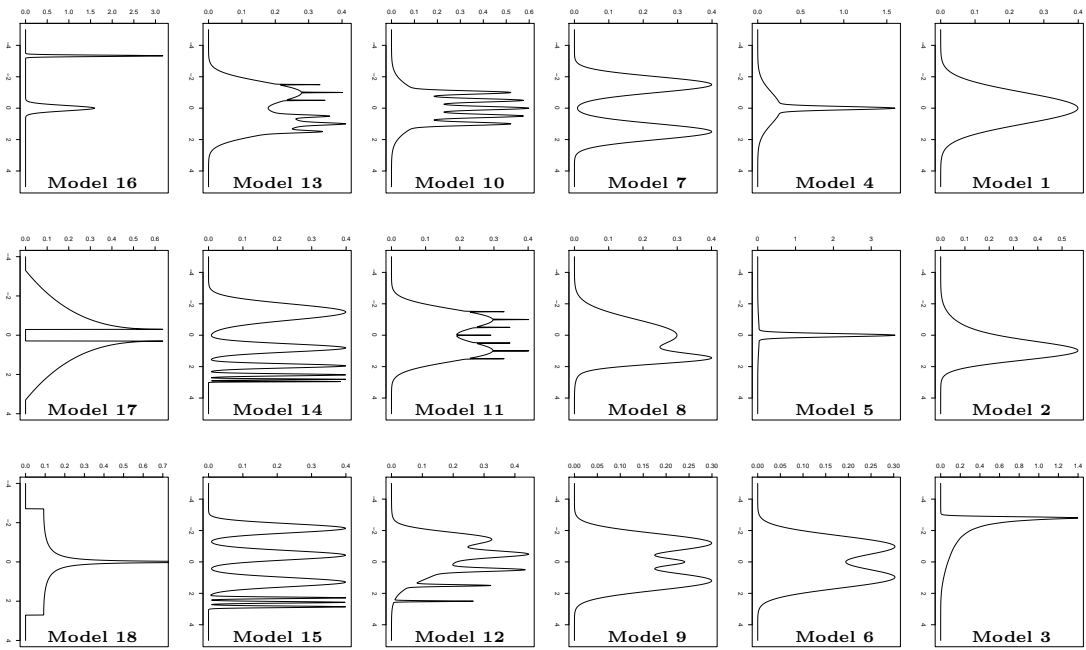

We generated $1000$ samples of size $n=1600$ from 18 density models (see Figure \ref{iok}). Models 1--15 are the densities proposed by Marron and Wand (1992). Models 16--18 correspond to the marronite, caliper, and matterhorn densities proposed in Berlinet and Devroye (1994). These models were added because they present some special properties. Specifically, they have two asymmetric and separated modes, two jumps, and a non-finite peak, respectively.

In addition, three values of the parameter $\tau$ have been considered: $\tau=0.2$, $\tau=0.5$, and $\tau=0.8$. The threshold $f_\tau$ was estimated using the two procedures described previously. Although the comparison is not shown, the two algorithms for estimating $f_\tau$ provided similar results. Therefore, in the following, only the results for Hyndman's method will be considered. The bandwidth given by Sheather and Jones has been used as the pilot selector to calculate $f_n$ for the estimation of $f_\tau$. %However, a new alternative is proposed in order to estimate $f_\tau$ for the Singh, Clayton and Nowak's algorithm. In this case, it will not be necessary to use a kernel density estimator for $f$. The threshold $f_\tau$ will be estimated empirically by imitating the procedure proposed by Walther (1997),
%$$\hat{f}_\tau =\max\left\{ \lambda>0:\frac{1}{n}\sum_{i=1}^n \mathbb{I}_{\{X_i\in\hat{L}(\lambda)\}}\geq 1-\tau  \right\}$$where $\hat{L}(\lambda)$ denotes the Singh, Clayton and Nowak's estimator for level $\lambda$.

\normalsize{

For each fixed random sample and each method, we calculated the estimator $\hat{L}(\tau)$ of $L(\tau)$ and the error in the estimation $d_{\mu_f}(L(\tau),\hat{L}(\tau))$. Thus, $1000$ errors were obtained for each algorithm, model, and value of $\tau$. To facilitate the presentation of the results, the following figures are divided into rectangles of different colours according to the method (vertical axis) and density model (horizontal axis), where light colours correspond to small errors, and vice versa. This representation allows us to detect the most competitive algorithm for each fixed value of $\tau$ and model. Given a density, the empirical means of the $1000$ errors have been ordered by testing whether the mean errors of the compared methods are equal. If the null hypothesis of equality between two methods is rejected, then each algorithm is painted a different colour (darker or lighter according to the mean error). Otherwise, both algorithms are represented using the same colour. This approach will be used in the rest of the paper.

}

}

\begin{figure}[h!]
\includegraphics[height=.45cm, width=.95\textwidth]{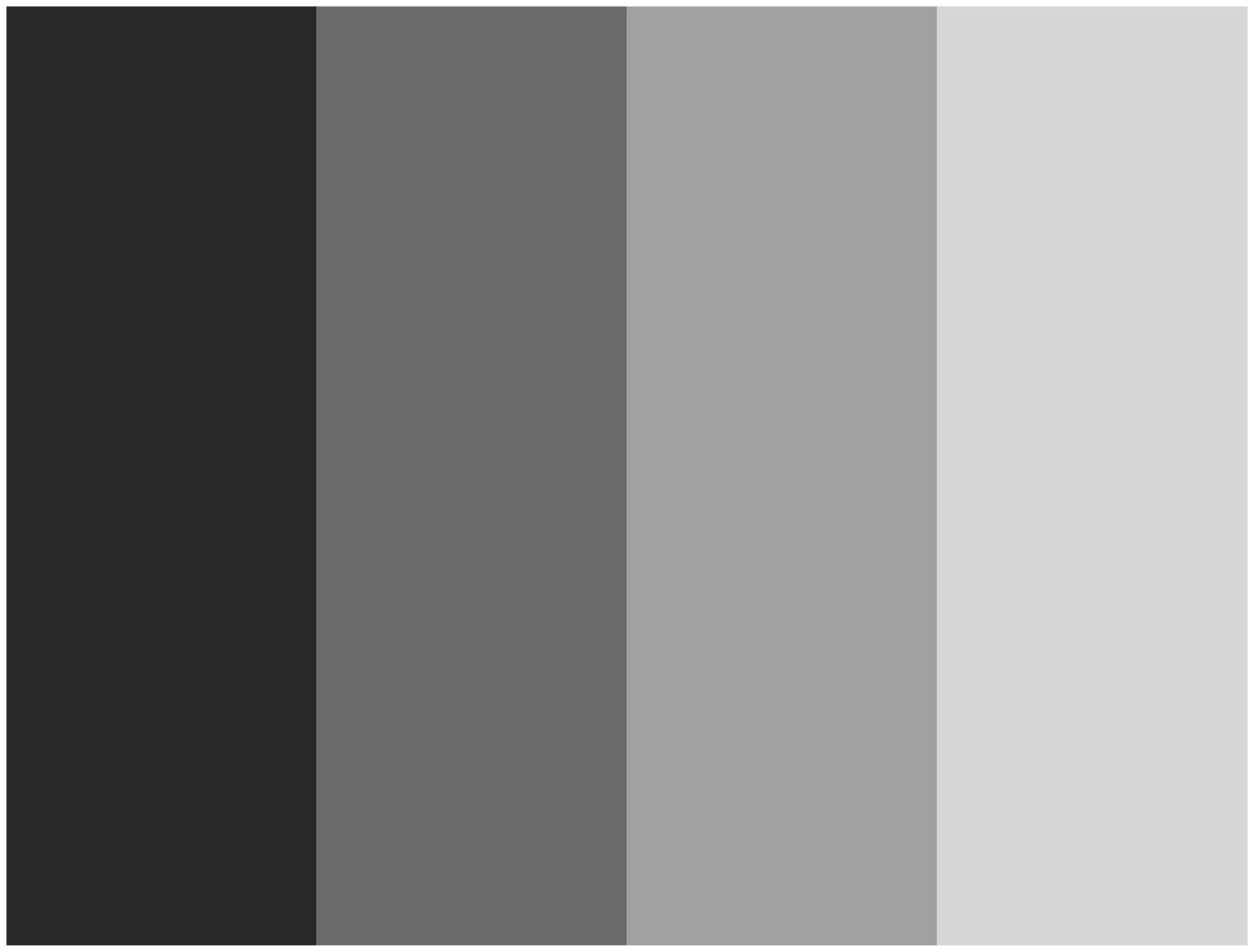}\vspace{-.75cm}\\
\hspace{-1.6cm}\includegraphics[height=8.2 cm, width=.95\textwidth]{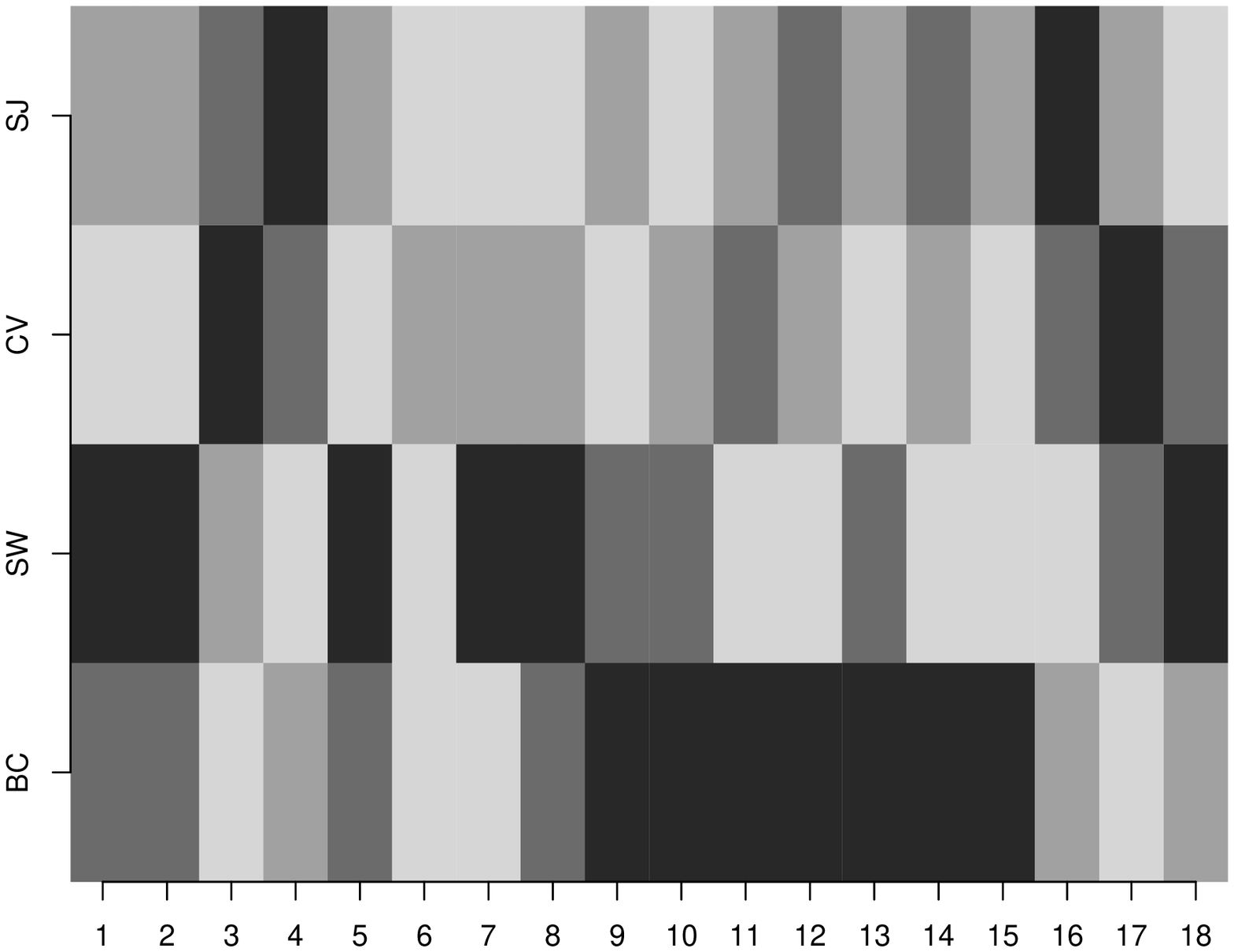}\vspace{-1.2 cm}\\
\caption{Comparison of plug-in methods (vertical axis) with the 18 model densities (horizontal axis), $\tau=0.2$ and $n=1600$.}\label{plugtau01}
\end{figure}

\begin{figure}[h!]
\includegraphics[height=.45cm, width=.95\textwidth]{Pluginleyenda.eps}\vspace{-.75cm}\\
\hspace{-1.6cm}\includegraphics[height=8.2 cm, width=.95\textwidth]{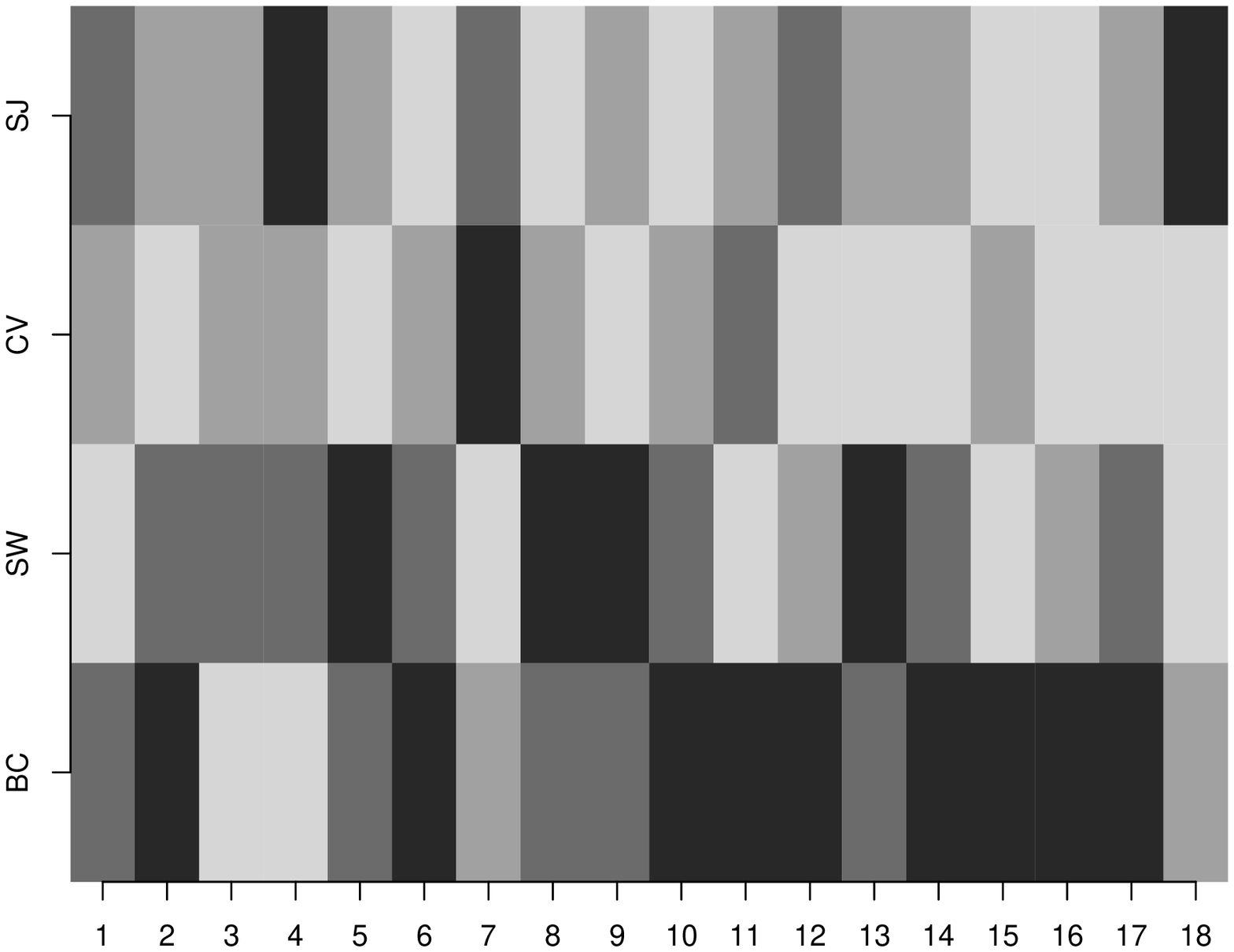}\vspace{-1.2 cm}\\
\caption{Comparison of plug-in methods (vertical axis) with the 18 model densities (horizontal axis), $\tau=0.5$ and $n=1600$.}\label{plugtau05}
\end{figure}

\begin{figure}[h!]
\includegraphics[height=.45cm, width=.95\textwidth]{Pluginleyenda.eps}\vspace{-.75cm}\\
\hspace{-1.6cm}\includegraphics[height=8.2 cm, width=.95\textwidth]{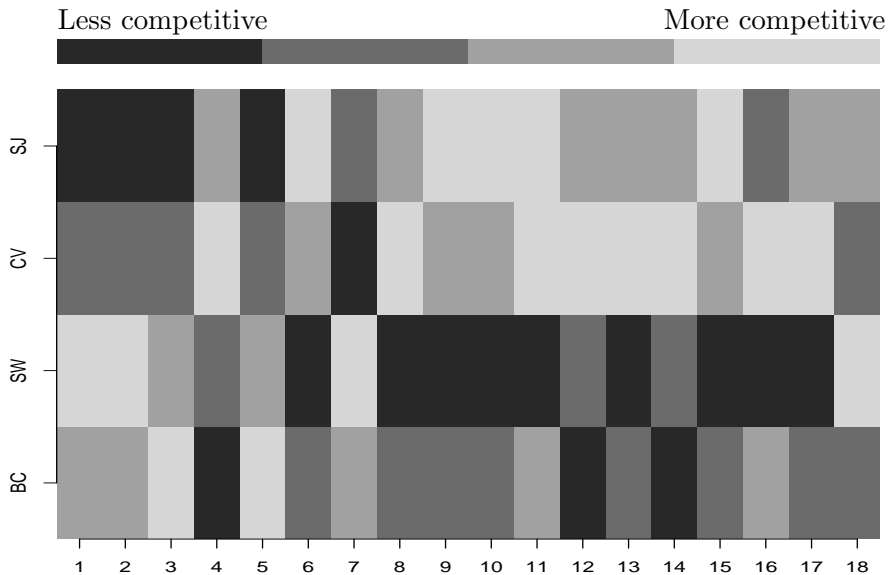}\vspace{-1.2 cm}\\
\caption{Comparison of plug-in methods (vertical axis) with the 18 model densities (horizontal axis), $\tau=0.8$ and $n=1600$.}\label{plugtau08}
\end{figure}

\normalsize{

Figures \ref{plugtau01}, \ref{plugtau05}, and \ref{plugtau08} compare the plug-in methods for $\tau=0.2$, $\tau=0.5$, and $\tau=0.8$, respectively. Clearly, the general bandwidth selectors exhibit the best global behaviour for $\tau=0.2$ and $\tau=0.5$. When $\tau=0.8$, the classical selectors again produce the best results for models 4, 6, 8, 9, 10, 11, 12, 13, 14, 15, 16, and 17. Some of these models are quite sophisticated densities. In this case, specific selectors are only competitive for quite simple models, such as 1, 2, 3, 5, 7, and 18 in Figure \ref{plugtau08}. All of these densities are unimodal or simple bimodal models.

In particular, if $\tau=0.2$ or $\tau=0.5$, SW behaves better than BC (see densities 10, 11, 12, 14, and 15 in Figures \ref{plugtau01} and \ref{plugtau05}). However, BC provides better results than SW for $\tau=0.8$ (see models 3, 5, 6, 7, 8, 9, 10, 11, 13, 15, 16, and 17 in Figure \ref{plugtau08}).

}

\newpage

\vspace{-.5cm}$ $ $ $ $ $ $  $ $ $ $ $ \small{Less competitive} $ $ $ $ $ $ $ $ $ $ $ $ $ $ $ $ $ $ $ $ $ $ $ $ $ $ $ $ $ $ $ $ $ $ $ $ $ $ $ $ $ $ $ $ $ $ $ $ $ $ $ $ $ $ $ $ $ $ $ $ $ $ $ $ $ $ $ $ $ $ $ $ $ $ $ $ $ $  \small{More competitive}\vspace{-.45cm}

 $ $

\vspace{.1cm}$ $ $ $ $ $ $  $ $ $ $ $ \small{Less competitive} $ $  $ $ $ $ $ $ $ $ $ $ $ $ $ $ $ $ $ $ $ $ $ $ $ $ $ $ $ $ $ $ $ $ $ $ $ $ $ $ $ $ $ $ $ $ $ $ $ $ $ $ $ $ $ $ $ $ $ $ $ $ $ $ $ $ $ $ $ $ $ $ $ $ $ $ $ $ $ $  \small{More competitive}\vspace{-.45cm}

\newpage

\vspace{-.5cm}$ $ $ $ $ $ $  $ $ $ $ $ \small{Less competitive} $ $  $ $ $ $ $ $ $ $ $ $ $ $ $ $ $ $ $ $ $ $ $ $ $ $ $ $ $ $ $ $ $ $ $ $ $ $ $ $ $ $ $ $ $ $ $ $ $ $ $ $ $ $ $ $ $ $ $ $ $ $ $ $ $ $ $ $ $ $ $ $ $ $ $ $ $ $ $ $  \small{More competitive}\vspace{-.45cm}

\normalsize{

In conclusion, specific methods to estimate level sets, such as those of Samworth and Wand, Ba\'illo and Cuevas, or Singh, Clayton and Nowak, do not improve the results of the classic
bandwidth selection rules, at least for the sample sizes considered here. In addition, the CV and SJ
methods often provide similar results, and present the best global behaviour.
}

%$ $ $ $ $ $ $ $ $ $ $ $ $ $ $ $ $ $ $ $ $ $ $ $ $ $\small{Less competitive}$ $ $ $ $ $ $ $ $ $ $ $ $ $ $ $ $ $ $ $ $ $ $ $ $ $ $ $ $ $ $ $ $ $ $ $ $ $ $ $ $ $ $ $ $ $ $ $ $ $ $ $ $ $ $ $ $ $ $ $ $ $ $ $ $ $ $ $ $ $ $ $ $ $ $ $ $ $ $ $ $ $ $ $ $ $ $ $ $ $ $ $ $ $ $ $ $ $ $ $ $ $ $ $ $ $ $ $ $ $ $ $ $ $ $ $ $ $ $ $ $ $ $ $ $ $ $ $ $ $ $ $ $ $ $ $ $ $ $ $  \small{More competitive}\vspace{-.55cm}
%\begin{figure}[h!]
%\includegraphics[height=.45cm, width=1.\textwidth]{Pluginleyenda.eps}\vspace{-.75cm}\\
%\includegraphics[height=7.65cm, width=1\textwidth]{Plugintau02n1600.eps}\vspace{-1 cm}\\
%\caption{Comparison of plug-in methods (vertical axis) with 15 density models (horizontal axis), $\tau=0.2$ and $n=1600$.}
%\end{figure}

\section{Comparative study of excess mass methods}\label{exceso de masa}

\normalsize{Another possibility is to assume that the set of interest satisfies some geometric condition, such as convexity. In this case, the excess mass approach, first proposed by Hartigan (1987) and M\"{u}ller and Sawitzki (1987), provides an alternative for the reconstruction of density level sets. Some previous contributions can be seen in  Chernoff (1964) or Eddy and Hartigan (1977). This group of algorithms utilizes the fact that the density level set $C(\lambda)=\{f\geq \lambda\}$, for a given value of $\lambda>0$, maximizes the functional\vspace{.1cm}
$$H_\lambda(B)=\mathbb{P}(B)-\lambda\mu(B),$$
where $\mathbb{P}$ is the probability measure induced by $f$ and $\mu$ is the Lebesgue measure. If $\mathcal{B}$ is a given class of sets, then a natural estimator $\hat{C}(\lambda)$ of $C(\lambda)$, under the shape restriction $C(\lambda)\in \mathcal{B}$, would be the maximizer of the empirical excess mass\vspace{.1cm}$$H_{\lambda,n}(B)=\mathbb{P}_n(B)-\lambda\mu(B)$$
on $ \mathcal{B}$, where $\mathbb{P}_n$ denotes the empirical probability induced by the sample $\mathcal{X}_n$. This method incorporates geometric information in a natural way. If no shape restriction is assumed, the maximizer of $H_{\lambda,n}$ would be $\mathcal{X}_n$. Hartigan
(1987) and Gr\"ubel (1988) considered the case where $ \mathcal{B}$ is the class of convex sets in the two- and one-dimensional cases, respectively. Asymptotic estimator results for more general classes
$ \mathcal{B}$ were given by Polonik (1995).

According to the previous comments, the goal of this work is to reconstruct density level sets with a fixed probability content. To avoid the bandwidth selection problem, and again following the empirical procedure proposed by Walther (1997), $\hat{L}(\tau)=\hat{C}(\hat{f}_\tau)$, where\vspace{.1cm}
$$\hat{f}_\tau =\max\{ \lambda>0: \mathbb{P}_n(\hat{C}(\lambda))\geq 1-\tau \}. $$

This methodology has been widely studied in the literature.
M\"uller and Sawitzki (1991) proposed an efficient algorithm for estimating one-dimensional sets by assuming that the theoretical level set can be written as a finite union of $M$ closed intervals. This algorithm assumes that $M$ is known a priori.

\subsection{Simulation results for excess mass methods}\label{Excessmasssimu}

\normalsize{

M\"{u}ller and Sawitzki's method depends on a parameter $M$, which, if unknown, presents the main disadvantage of this algorithm. Five values for the number of clusters have been considered, $M=1,\mbox{ }2,\mbox{ }3,\mbox{ }4,$ and $5$. The M\"{u}ller and Sawitzki's method with $M$ modes will be denoted by MS$_M$.

Figure \ref{muller} shows the simulation results for the five values of $M$ with $\tau=0.5$ and $n=1600$. These can be used to analyze the influence of $M$ in the method of  M\"{u}ller and Sawitzki. The real number of modes for each density when $\tau=0.5$ has been included at the top of the vertical axis. It is clear that M\"uller and Sawitzki's algorithm is very sensitive to the parameter $M$. For $\tau=0.5$, densities 1--5 are unimodal, and MS$_1$ provides the best results. Densities 6--8 have two modes and, in this case, the best algorithm is MS$_2$. However, for the two modes of model 9, MS$_3$ provides the best results. Model 10 has five modes for $\tau=0.5$, and here MS$_5$ produces the best estimations. However, the optimum value of $M$ for M\"uller and Sawitzki's method is not equal to the real value of $M$ for models 11, 12, and 13, because some of their modes are not significant. In addition, if $M$ is misspecified, larger values of $M$ tend to be better than smaller ones (see models 6--9).
}

$\vspace{.1cm} $

$ $ $ $ $ $ $ $ $ $ $ $ \small{Less competitive}  $ $ $ $ $ $ $ $ $ $ $ $ $ $ $ $ $ $ $ $ $ $ $ $ $ $ $ $ $ $ $ $ $ $ $ $ $ $ $ $ $ $ $ $ $ $ $ $ $ $ $ $ $ $ $ $ $ $ $ $ $ $  $ $ $ $ $ $ $ $ $ $ $ $ $ $ $ $ \small{More competitive}\vspace{-.4cm}
\begin{figure}[h!]
\includegraphics[height=.45cm, width=.95\textwidth]{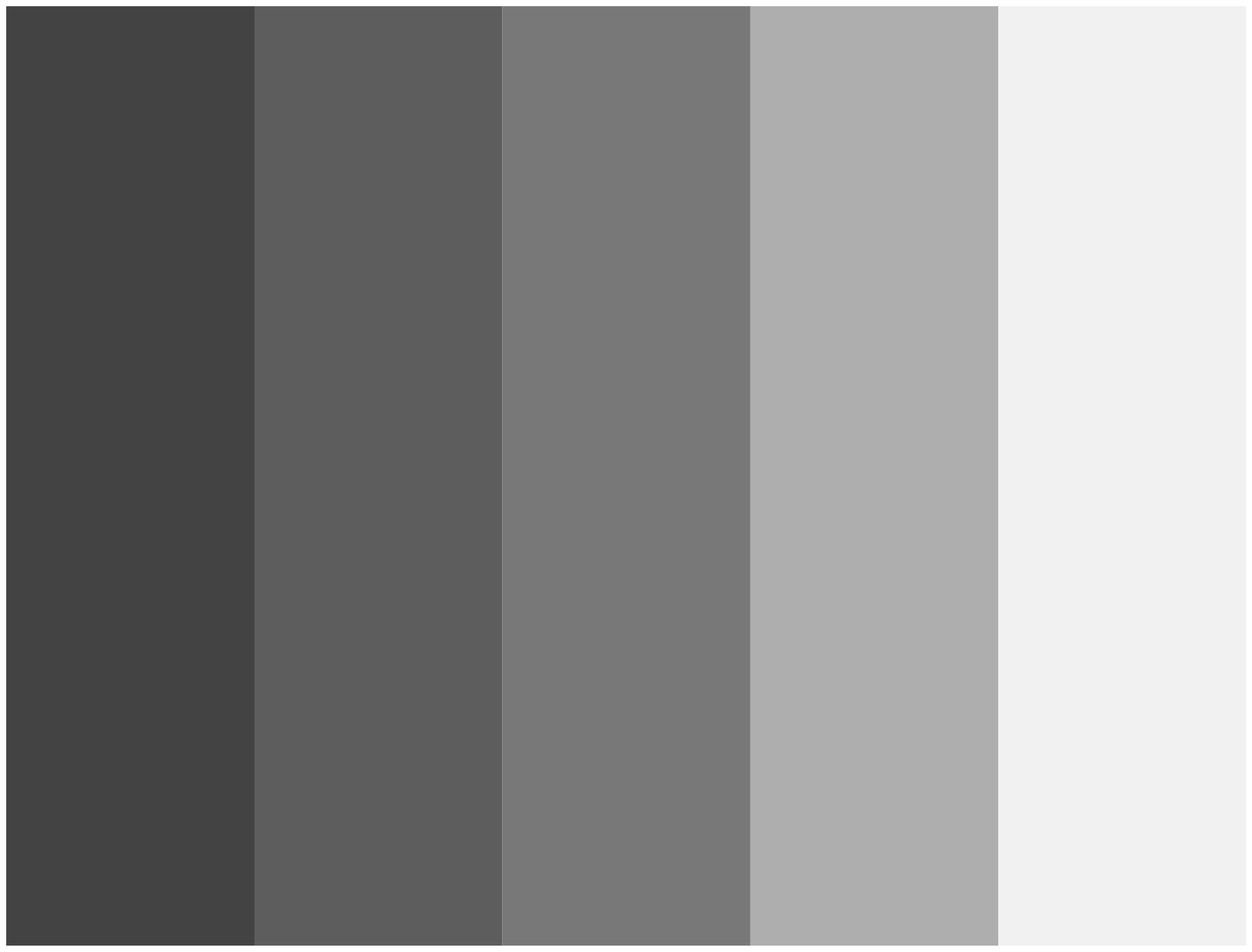}\vspace{-1.3cm}\\
\hspace{-1.6cm}\includegraphics[height=9.5 cm, width=.95\textwidth]{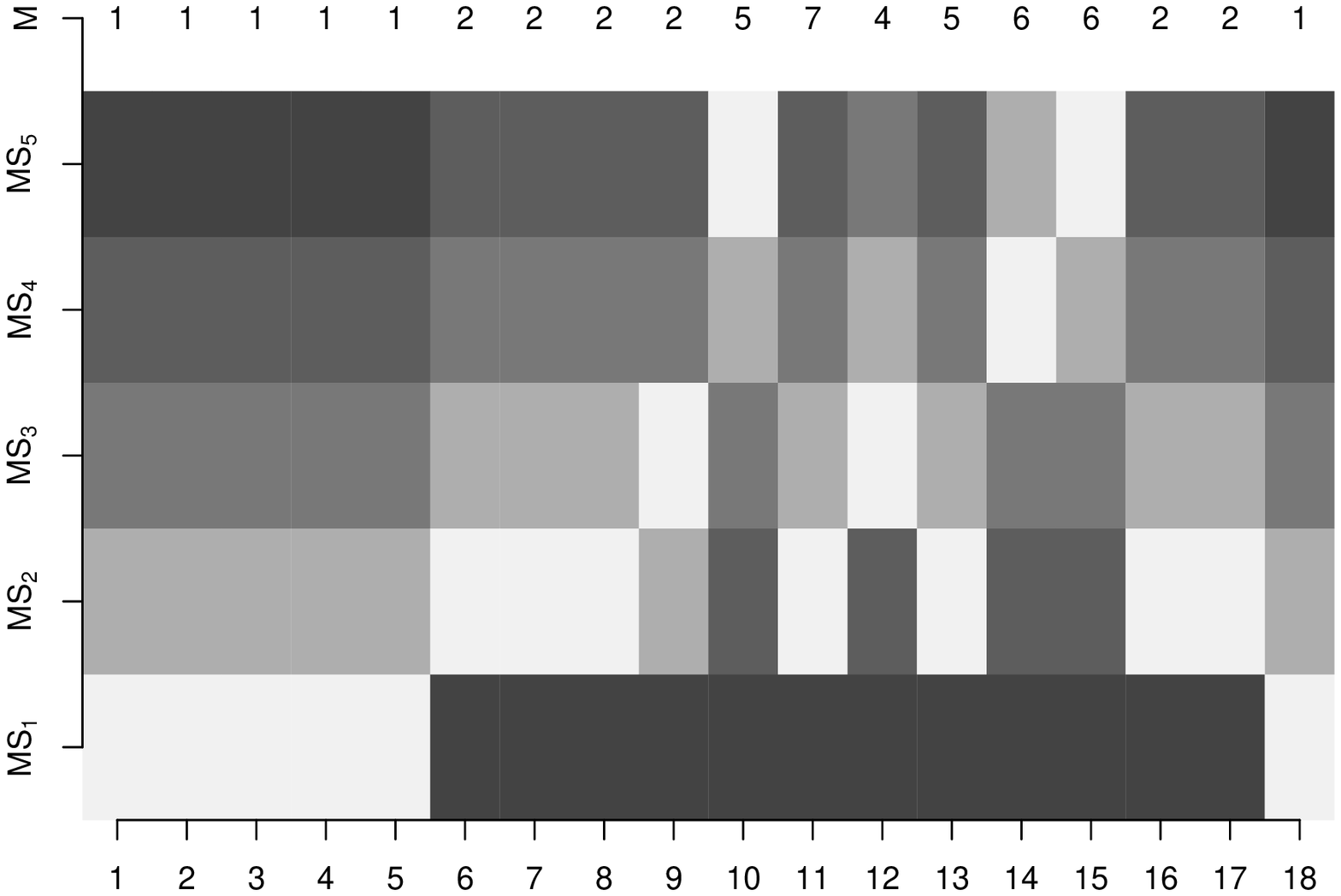}\vspace{-1.3cm}\\
\caption{Comparison of M\"{u}ller and Sawitzki's method for different values of $M$ (vertical axis) with the 18 model densities (horizontal axis), $\tau=0.5$ and $n=1600$. The real number of modes for each density has been included at the top of the vertical axis.}\label{muller}\vspace{-0.1cm}
\end{figure}

\normalsize{
%Again, we have tested if the means of errors are or not equal. If we reject the null hypothesis of equality between two means for the same model then each method will be painted a different color (darker or lighter according to the mean of the errors is higher or lower). In another case, both algorithms are represented using the same color.}\vspace{.3cm}

}

\section{Comparative study of hybrid methods}\label{hybridos}

\normalsize{

As the name suggests, hybrid methods assume a priori geometric restrictions on $L(\tau)$, and they also use a pilot nonparametric density estimator to define the set $\mathcal{X}_n^+=\{X_i\in\mathcal{X}_n:f_n(X_i)\geq \hat{f}_\tau\}$. In this paper, two new hybrid methods to estimate convex and $r$-convex ($r>0$) sets are proposed. This latter shape restriction generalizes the convexity assumption. A closed set $A$ is said to be $r$-convex for some $r>0$ if $A=C_{r}(A)$, where $$C_{r}(A)=\bigcap_{\{B_r(x):B_r(x)\cap
A=\emptyset\}}\left(B_r(x)\right)^c$$
denotes the $r$-convex hull
of $A$, $B_r(x)$ is the open ball with centre $x$ and radius $r$, and $\left(B_r(x)\right)^c$ is its complement.
Our two new proposals are based on the convex hull and $r$-convex hull methods for estimating the support (see Korostel\"ev and Tsybakov (1993) and Rodr\'iguez-Casal (2007), respectively). Under the convexity restriction, it is shown that we can estimate the level set as the convex hull of $\mathcal{X}_n^+$ and, under the $r$-convexity assumption, as the $r$-convex hull of $\mathcal{X}_n^+$.
Another classic hybrid technique is the granulometric smoothing method (see Walther, 1997). This assumes that the level set $L(\tau)$ and its complement are both $r$-convex. This method adapts  the support estimator proposed by Devroye and Wise (1980) to the context of level set estimation. In this case, the estimator is defined as the union of balls of radius $r$ around those points in $\mathcal{X}_n^+$ that
are a distance of at least $r$ from each point in $\mathcal{X}_n\setminus\mathcal{X}_n^+$.

}

\subsection{Simulation results for hybrid methods}\label{hibridossimu}

\normalsize{

Granulometric smoothing and the $r$-convex hull method depend on the unknown parameter $r$. This is the main disadvantage of these algorithms. In this work, five fixed values for the radius have been considered: $r_1 = 0.01$, $r_2 = 0.05$, $r_3 = 0.1,$ $r_4 = 0.2$, and $r_5 = 0.3$. Although not discussed here, the influence of the parameter $r$ on the $r$-convex hull method and granulometric smoothing has been studied, and in general, the $r$-convex hull method is less sensitive to the selection of $r$.

We denote the convex hull method as CH, the $r$-convex hull method as CH$_r$, and granulometric smoothing with radius $r$ as W$_r$.
The latter two methods have been compared by fixing $r=r_3$. For the first two, the threshold was calculated using Hyndman's method by taking Sheather and Jones' bandwidth as the pilot selector to calculate $f_n$. For W$_r$, the threshold was estimated empirically according to Walther's proposal.

Figures \ref{hibridos1}, \ref{hibridos2}, and \ref{hibridos3} show the results obtained for $\tau=0.2$, $\tau=0.5$, and $\tau=0.8$, respectively. Each method is represented on the vertical axis, and each density model on the horizontal axis.

Some density models present convex level sets for $\tau=0.2$, although they are not unimodal (for example, densities 6, 8, or 11). In these cases, when the convexity assumption is true, CH can be very competitive. However, models 1--4 have convex level sets for any value of $\tau$, and the $r_3$-convex hull method is then the most competitive. In addition, sometimes the convexity hypothesis can be very restrictive (models 7 or 10, for example), and in such cases the $r_3$-convex hull method or granulometric smoothing provide better results.

}

 \newpage

$ $ $ $ $ $ $ $ $ $ $ $\small{Less competitive}   $ $ $ $ $ $ $ $ $ $ $ $ $ $ $ $ $ $ $ $ $ $ $ $ $ $ $ $ $ $ $ $ $ $ $ $ $ $ $ $ $ $ $ $ $ $ $ $ $ $ $ $ $ $ $ $ $ $ $ $ $ $ $ $ $ $ $ $ $ $  $ $ $ $ $ $\small{More competitive}\vspace{-.45cm}
\begin{figure}[h!]
\includegraphics[height=.45cm, width=.92\textwidth]{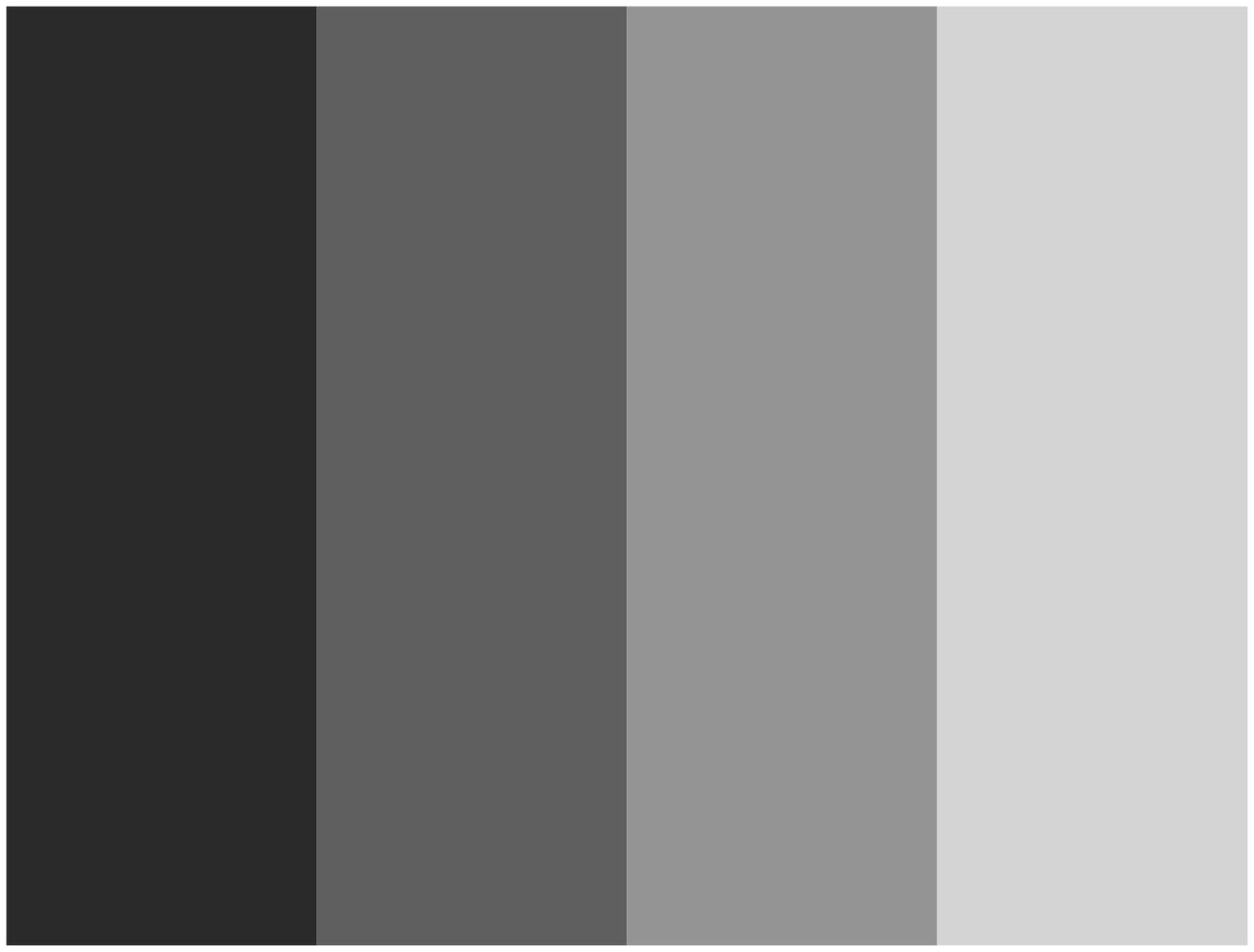}\vspace{-.75cm}\\
\hspace{-1.6cm}\includegraphics[height=8.4cm, width=.92\textwidth]{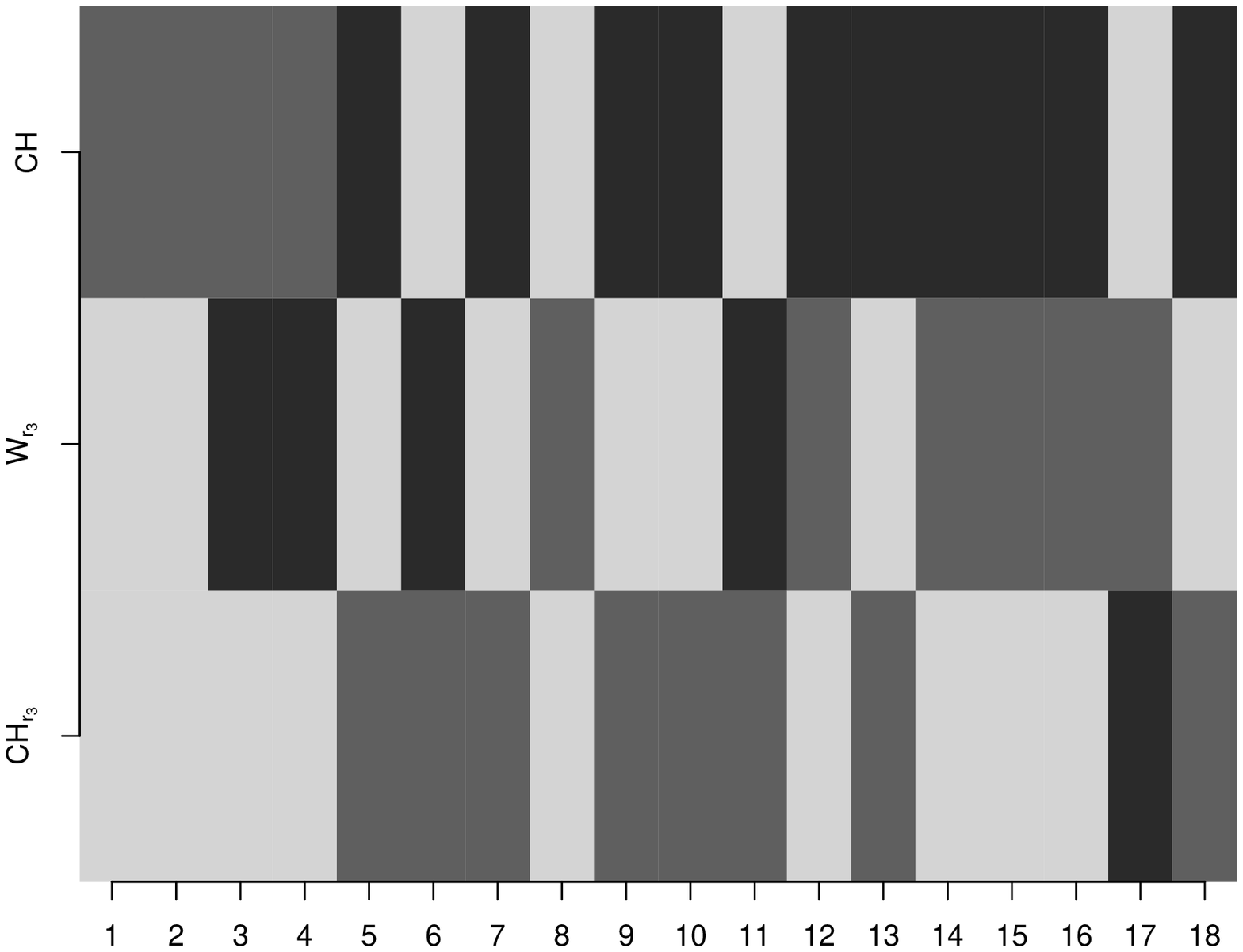}\vspace{-1.1cm}\\
\caption{Comparison of hybrid methods (vertical axis) with the 18 model densities (horizontal axis), $\tau=0.2$ and $n=1600$.}\label{hibridos1}
\end{figure}

\vspace{.2cm}$ $ $ $ $ $ $ $ $ $ $ $\small{Less competitive}   $ $ $ $ $ $ $ $ $ $ $ $ $ $ $ $ $ $ $ $ $ $ $ $ $ $ $ $ $ $ $ $ $ $ $ $ $ $ $ $ $ $ $ $ $ $ $ $ $ $ $ $ $ $ $ $ $ $ $ $ $ $ $ $ $ $ $ $ $ $  $ $ $ $ $ $\small{More competitive}\vspace{-.45cm}
\begin{figure}[h!]
\includegraphics[height=.45cm, width=0.92\textwidth]{HYleyenda.eps}\vspace{-.75cm}\\
\hspace{-1.6cm}\includegraphics[height=8.4cm, width=0.92\textwidth]{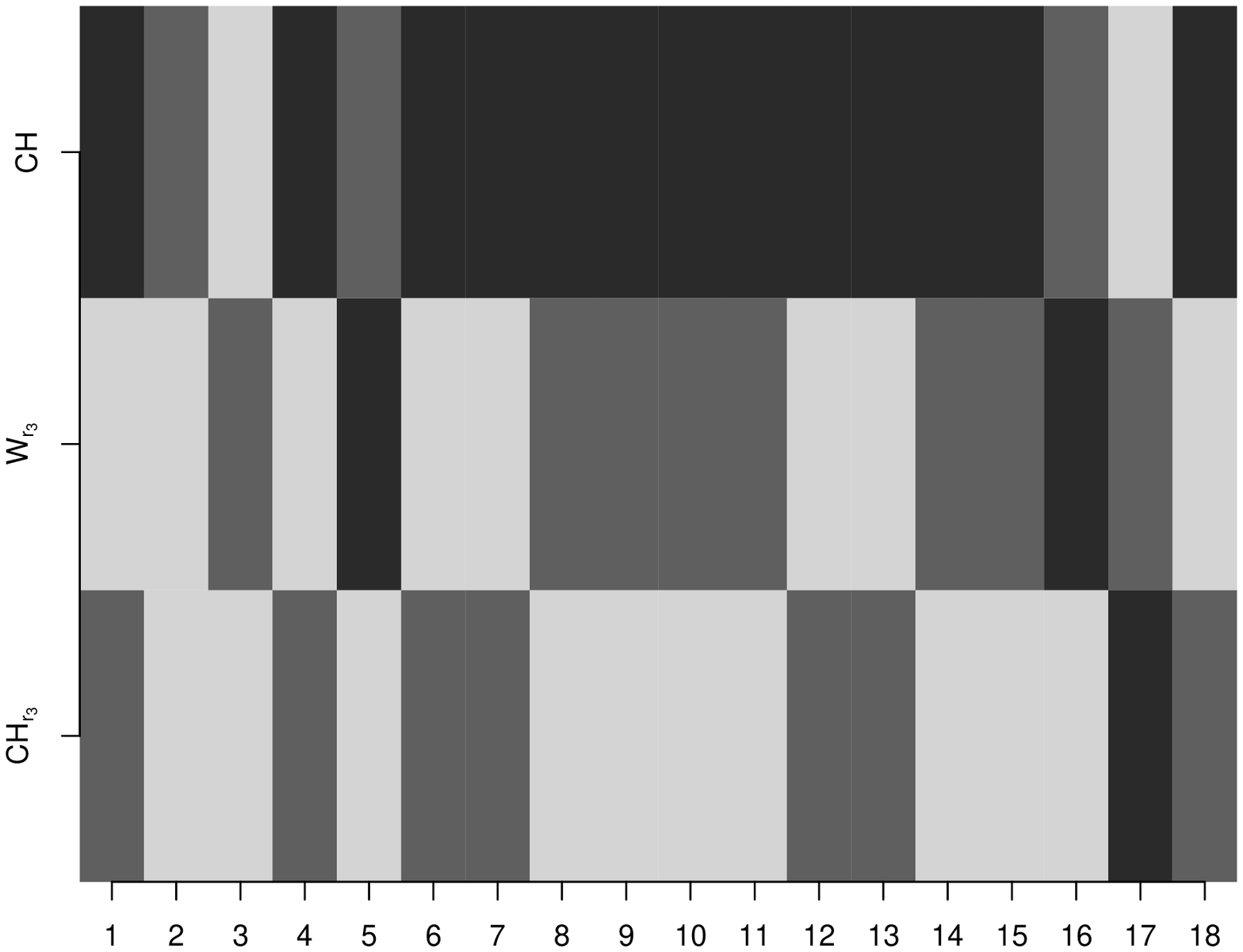}\vspace{-1.25cm}\\
\caption{Comparison of hybrid methods (vertical axis) with the 18 model densities (horizontal axis), $\tau=0.5$ and $n=1600$.}\label{hibridos2}
\end{figure}

 \newpage

$ $ $ $ $ $ $ $ $ $ $ $\small{Less competitive}   $ $ $ $ $ $ $ $ $ $ $ $ $ $ $ $ $ $ $ $ $ $ $ $ $ $ $ $ $ $ $ $ $ $ $ $ $ $ $ $ $ $ $ $ $ $ $ $ $ $ $ $ $ $ $ $ $ $ $ $ $ $ $ $ $ $ $ $ $ $  $ $ $ $ $ $\small{More competitive}\vspace{-.45cm}
\begin{figure}[h!]
\includegraphics[height=.45cm, width=0.92\textwidth]{HYleyenda.eps}\vspace{-.75cm}\\
\hspace{-1.6cm}\includegraphics[height=8.4cm, width=0.92\textwidth]{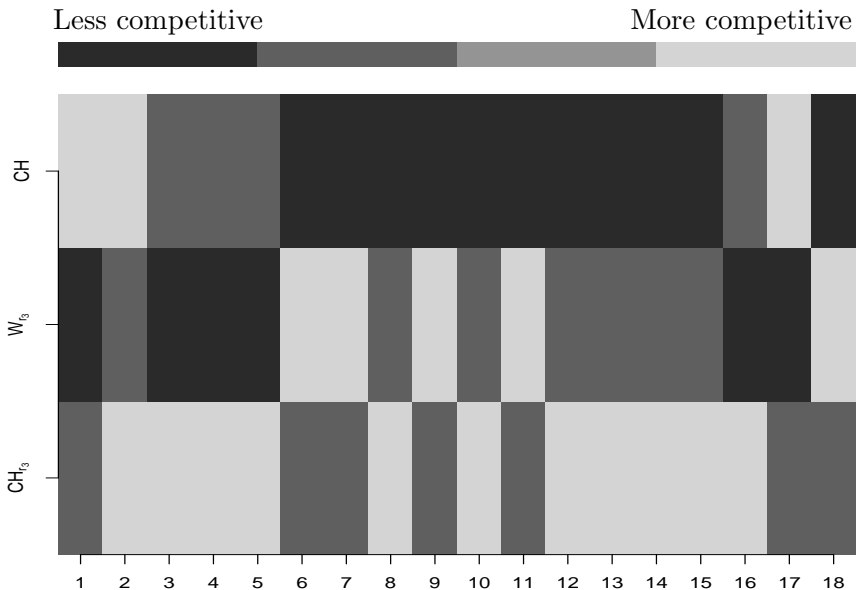}\vspace{-1.25cm}\\
\caption{Comparison of hybrid methods (vertical axis) with the 18 model densities (horizontal axis), $\tau=0.8$ and $n=1600$.}\label{hibridos3}
\end{figure}

%\newpage
%$ $ $ $ $ $ $ $  $ $ $ $ $ $ $ $ $ $ $ $ $ $ $ $ $ $ $ $ $ $ $ $ $ $\small{Less competitive}  $ $ $ $ $ $ $ $ $ $ $ $ $ $ $ $ $ $ $ $ $ $ $ $ $ $ $ $ $ $ $ $ $ $ $ $ $ $ $ $ $ $ $ $ $ $ $ $ $ $ $ $ $ $ $ $ $ $ $ $ $ $ $ $ $ $ $ $ $ $ $ $ $ $ $ $ $ $ $ $ $ $ $ $ $ $ $ $ $ $ $ $ $ $ $ $ $ $ $ $ $ $ $ $ $ $ $ $ $ $ $ $ $ $  $ $ $ $ $ $\small{More competitive}\vspace{-.55cm}
%\begin{figure}[h!]
%\includegraphics[height=.45cm, width=1.\textwidth]{HYleyenda.eps}\vspace{-.75cm}\\
%\hspace{-1.6cm}\includegraphics[height=7.65cm, width=1.\textwidth]{hytau05n1600.eps}\vspace{-1cm}\\
%\caption{Comparison of hybrid methods (vertical axis) with 15 density models (horizontal axis), $\tau=0.5$ and $n=1600$.}
%\end{figure}

\normalsize{

}

\normalsize{
When  $\tau=0.5$, convexity is a very restrictive shape condition for most of the models (see densities 6--15). Granulometric smoothing and the $r_3$-convex hull method have better behaviour than CH in this case. The $r_3$-convex hull algorithm is quite competitive for multimodal models, such as for densities 8, 9, 10, 11, 14, and 15.
}

\normalsize{

Although Walther's method and the $r_3$-convex hull algorithm provide similar results for $\tau=0.2$ and $\tau=0.5$, the new proposal is clearly better for high values of $\tau$---see unimodal models 2, 3, 4, 5, and 8 or more complex densities 10, 12, 13, 14, 15, and 16 in Figure \ref{hibridos3}. Granulometric smoothing is only the most competitive method for densities 6, 7, 9, 11, and 18. The first three have quite simple level sets, with only two connected components, whereas model 11 presents seven modes, five of which are not significant.

Note that the $r_3$-convex hull method does not provide the worst result for any of the models considered. This is quite promising, because the parameter $r$ is not estimated from the data, and it was expected that this would change from model to model.

}

\section{Final comparison}\label{comparacionfinal}
\normalsize{
Finally, we compared the most competitive methods in each group: cross-validation, M\"uller and Sawitzki's method, granulometric smoothing, and the $r$-convex hull method. It is necessary to specify values for $M$ and $r$ for M\"uller and Sawitzki's method, granulometric smoothing, and the $r$-convex hull method. We therefore considered $M=1$, $M=2$, and $r=r_3$.

Figures \ref{final1}, \ref{final2}, and \ref{final12} show the results for $\tau=0.2$, $\tau=0.5$, and $\tau=0.8$, respectively. When $\tau=0.2$, M\"uller and Sawitzki's method is more competitive with $M=1$ than with $M=2$ for unimodal densities of 1--5, as shown in Figure \ref{final1}. The same conclusion can be extracted for models 6, 8, 11, and 18, because, for this value of $\tau$, their level sets have only one interval. M\"uller and Sawitzki's method with $M=2$ does not exhibit very good results, because most of the models are not bimodal. In spite of this, it is the most competitive algorithm for models  10, 12, and 16. In this case, cross-validation provides quite good results, except for densities 3, 4, 6, 7, 9, 11, and 17, where one of the two hybrid methods presents the best behaviour. However, these two algorithms have a very important disadvantage in that they depend on an unknown parameter.

}

\normalsize{

When $\tau=0.5$, the cross-validation selector provides quite competitive results (see models 1, 2, 4, 5, 6, 10, 12, 14, 16, 17, and 18 in Figure \ref{final2}). Granulometric smoothing and the $r_3$-convex hull method produce the best results for densities 6, 7, and 8 and 3, 8, 9, 11, 15, and 16, respectively. In this case, M\"uller and Sawitzki's method is not particularly competitive, especially when $M=1$ for models 6--17. All of these densities have level sets with more than one interval.

}

\normalsize{

Cross-validation exhibits the most competitive behaviour for unimodal densities when $\tau=0.8$ (see models 1, 2, 4, 5, 8, and  density 16 in Figure \ref{final12}). Granulometric smoothing presents its worst performance for densities 3, 4, 5, and 16. However, although the $r_3$-convex hull method is not the most competitive for many of the models, it presents very regular behaviour.  M\"uller and Sawitzki's method with $M=1$ does not provide very good results, e.g., densities 6, 7, 9, 10, 11, 12, 14, and 15. None of these models are unimodal. If we consider $M=2$, this method is the most competitive for densities 12, 13, and 17.

}
\newpage

\vspace{-.95cm}

%$ $ $ $ $ $ $ $ $ $ $ $ \small{Less competitive}$ $ $ $ $ $ $  $ $ $ $ $ $ $ $ $ $ $ $ $ $ $ $ $ $ $ $ $ $ $ $ $ $ $ $ $ $ $ $ $ $ $ $ $ $ $ $ $ $ $ $ $ $ $ $ $ $ $ $ $ $ $  $ $ $ $ $ $ $ $ $ $ $ $ $ $ $ $ \small{More competitive}\vspace{-.4 cm}
%\begin{figure}[h!]
%\includegraphics[height=.4cm, width=0.94\textwidth]{Fleyenda.eps}\vspace{-.78cm}\\
%\hspace{-3.9 cm}\includegraphics[height=8.5cm, width=0.94\textwidth]{Ftau02n1600.eps}\vspace{-1.25 cm}\\
%\caption{Comparison of the CV, W$_{r_3}$, CH$_{r_3}$ and MS$_2$ (vertical axis) with the 18 model densities (horizontal axis), $\tau=0.2$ and $n=1600$.}\label{final1}
%\end{figure}% SE QUEREMOS QUITAR O MS1 quita % e listo

$ $ $ $ $ $ $ $ $ $ $ $ \small{Less competitive}$ $ $ $ $ $ $  $ $ $ $ $ $ $ $ $ $ $ $ $ $ $ $ $ $ $ $ $ $ $ $ $ $ $ $ $ $ $ $ $ $ $ $ $ $ $ $ $ $ $ $ $ $ $ $ $ $ $ $ $ $ $  $ $ $ $ $ $ $ $ $ $ $ $ $ $ $ $ \small{More competitive}\vspace{-.4 cm}
\begin{figure}[h!]
\includegraphics[height=.4cm, width=0.94\textwidth]{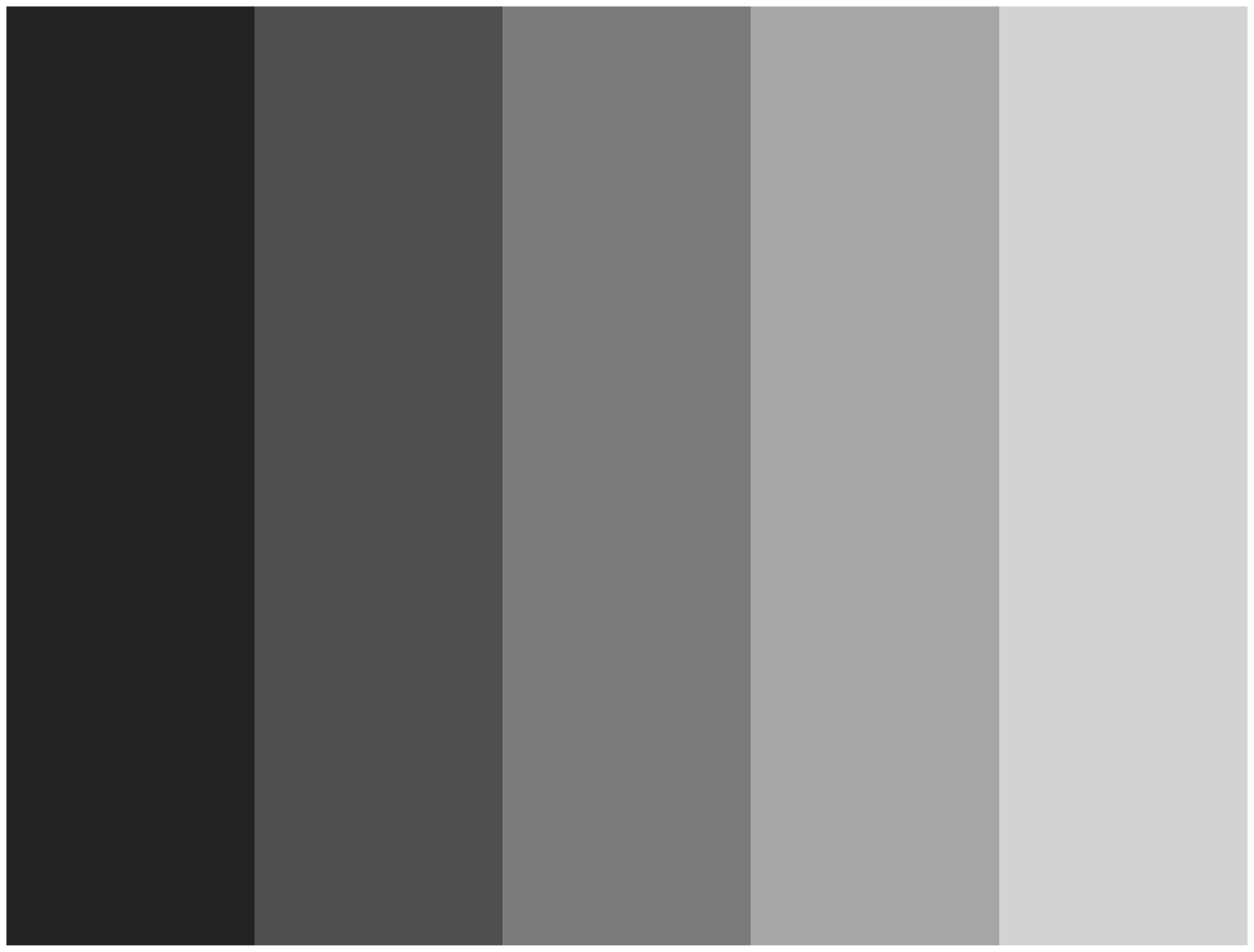}\vspace{-.78cm}\\
\hspace{-3.9 cm}\includegraphics[height=8.5cm, width=0.94\textwidth]{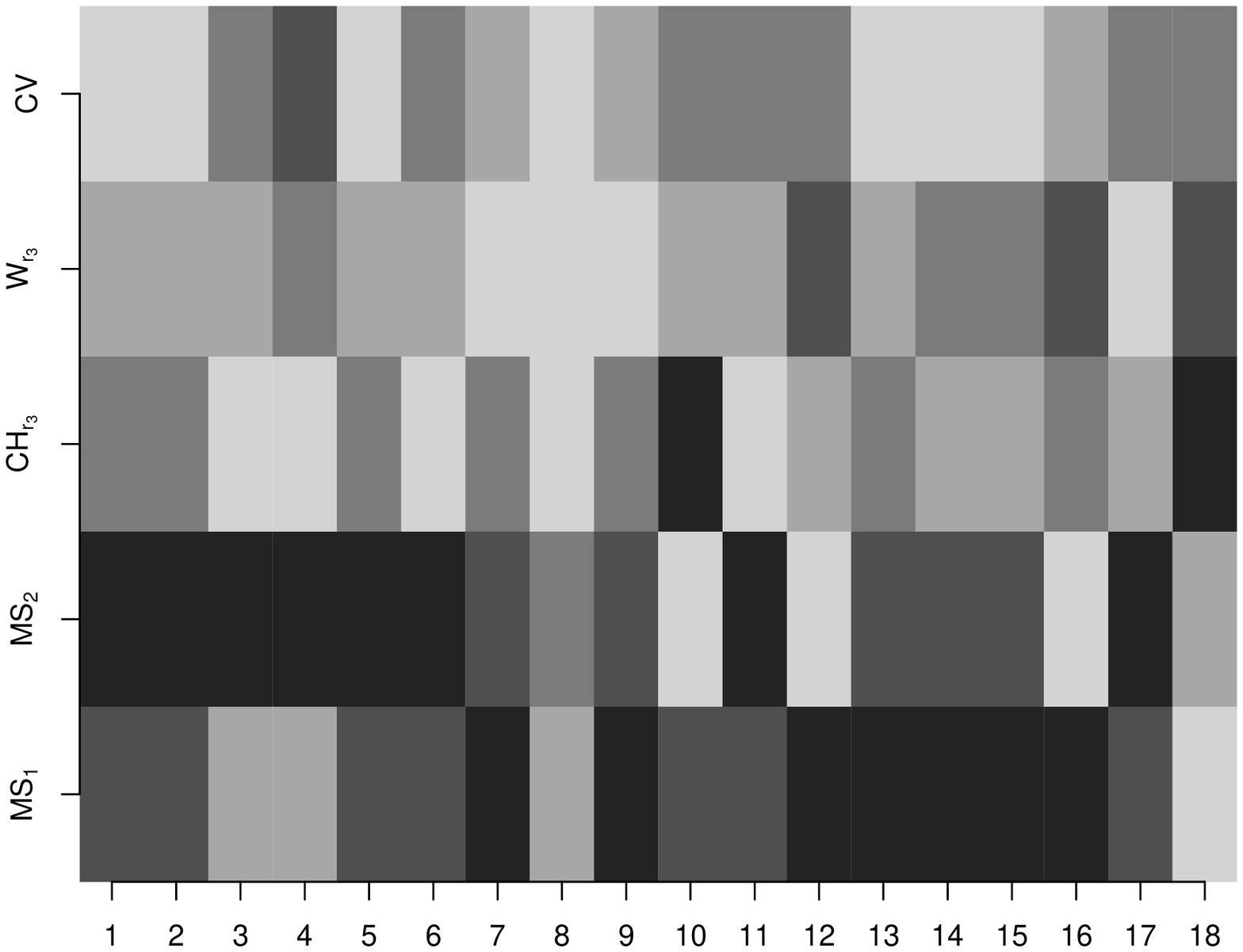}\vspace{-1.25 cm}\\
\caption{Comparison of CV, W$_{r_3}$, CH$_{r_3}$, MS$_2$, and MS$_1$ (vertical axis) with the 18 model densities (horizontal axis), $\tau=0.2$ and $n=1600$.}\label{final1}
\end{figure}

 \vspace{.2cm}

%$ $ $ $ $ $ $ $ $ $ $ $ \small{Less competitive}$ $ $ $ $ $ $  $ $ $ $ $ $ $ $ $ $ $ $ $ $ $ $ $ $ $ $ $ $ $ $ $ $ $ $ $ $ $ $ $ $ $ $ $ $ $ $ $ $ $ $ $ $ $ $ $ $ $ $ $ $ $  $ $ $ $ $ $ $ $ $ $ $ $ $ $ $ $ \small{More competitive}\vspace{-.4 cm}
%\begin{figure}[h!]
%\includegraphics[height=.4cm, width=.94\textwidth]{Fleyenda.eps}\vspace{-.78cm}\\
%\hspace{-3.9 cm}\includegraphics[height=8.5cm, width=.94\textwidth]{Ftau05n1600.eps}\vspace{-1.25 cm}\\
%\caption{Comparison of the CV, W$_{r_3}$, CH$_{r_3}$ and MS$_2$ (vertical axis) with the 18 model densities (horizontal axis), $\tau=0.5$ and $n=1600$.}\label{final2}
%\end{figure}

$ $ $ $ $ $ $ $ $ $ $ $ \small{Less competitive}$ $ $ $ $ $ $  $ $ $ $ $ $ $ $ $ $ $ $ $ $ $ $ $ $ $ $ $ $ $ $ $ $ $ $ $ $ $ $ $ $ $ $ $ $ $ $ $ $ $ $ $ $ $ $ $ $ $ $ $ $ $  $ $ $ $ $ $ $ $ $ $ $ $ $ $ $ $ \small{More competitive}\vspace{-.4 cm}
\begin{figure}[h!]
\includegraphics[height=.4cm, width=.94\textwidth]{Fleyenda2.eps}\vspace{-.78cm}\\
\hspace{-3.9 cm}\includegraphics[height=8.5cm, width=.94\textwidth]{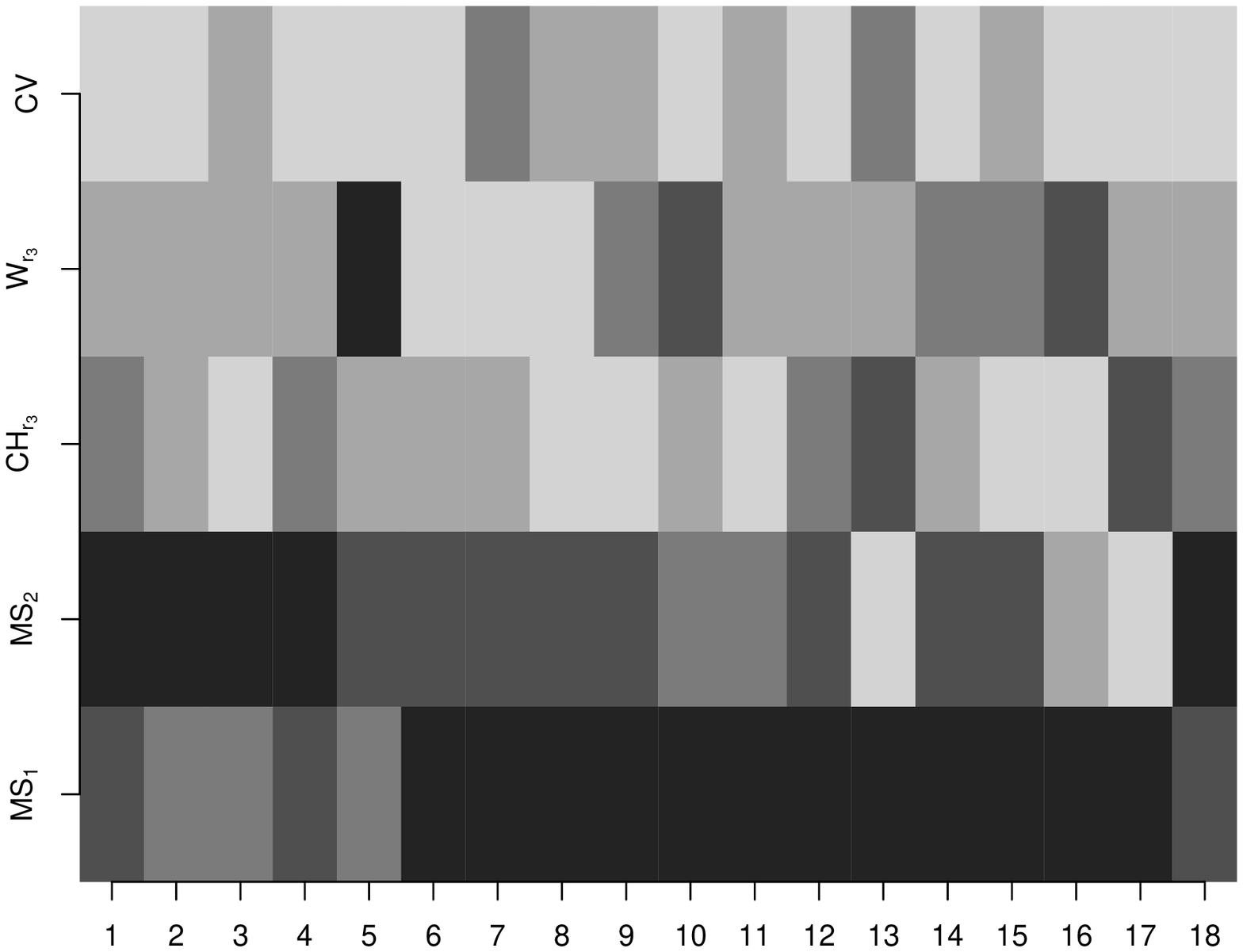}\vspace{-1.25 cm}\\
\caption{Comparison of CV, W$_{r_3}$, CH$_{r_3}$, MS$_2$, and MS$_1$ (vertical axis) with the 18 model densities (horizontal axis), $\tau=0.5$ and $n=1600$.}\label{final2}
\end{figure}

\newpage

$\vspace{-.9cm} $

%$ $ $ $ $ $ $ $ $ $ $ $ \small{Less competitive}$ $ $ $ $ $ $  $ $ $ $ $ $ $ $ $ $ $ $ $ $ $ $ $ $ $ $ $ $ $ $ $ $ $ $ $ $ $ $ $ $ $ $ $ $ $ $ $ $ $ $ $ $ $ $ $ $ $ $ $ $ $  $ $ $ $ $ $ $ $ $ $ $ $ $ $ $ $ \small{More competitive}\vspace{-.4 cm}
%\begin{figure}[h!]
%\includegraphics[height=.4cm, width=.94\textwidth]{Fleyenda.eps}\vspace{-.78cm}\\
%\hspace{-3.9 cm}\includegraphics[height=8.5cm, width=.94\textwidth]{Ftau08n1600.eps}\vspace{-1.25cm}\\
%\caption{Comparison of the CV, W$_{r_3}$, CH$_{r_3}$ and MS$_2$ (vertical axis) with the 18 model densities (horizontal axis), $\tau=0.8$ and $n=1600$.}\label{final12}
%\end{figure}

$ $ $ $ $ $ $ $ $ $ $ $ \small{Less competitive}$ $ $ $ $ $ $  $ $ $ $ $ $ $ $ $ $ $ $ $ $ $ $ $ $ $ $ $ $ $ $ $ $ $ $ $ $ $ $ $ $ $ $ $ $ $ $ $ $ $ $ $ $ $ $ $ $ $ $ $ $ $  $ $ $ $ $ $ $ $ $ $ $ $ $ $ $ $ \small{More competitive}\vspace{-.4 cm}
\begin{figure}[h!]
\includegraphics[height=.4cm, width=.94\textwidth]{Fleyenda2.eps}\vspace{-.78cm}\\
\hspace{-3.9 cm}\includegraphics[height=8.5cm, width=.94\textwidth]{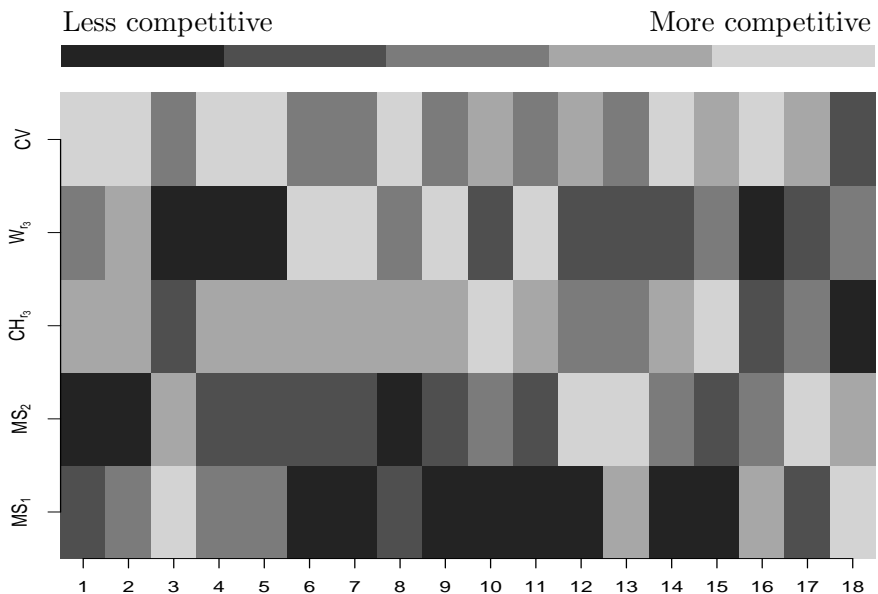}\vspace{-1.25cm}\\
\caption{Comparison of CV, W$_{r_3}$, CH$_{r_3}$, MS$_2$, and MS$_1$ (vertical axis) with the 18 model densities (horizontal axis), $\tau=0.8$ and $n=1600$.}\label{final12}
\end{figure}

\section{Conclusions}\label{conc}

\normalsize{

As has been stated previously, if no assumption is made on the shape of the density level set to be estimated, then plug-in methods provide good results. In general, cross-validation or even Sheather and Jones' method are good alternatives for reconstructing the level set. Previous results suggest that specific bandwidth selectors for density level sets present worse general behaviour for the sample size considered.

In contrast, excess mass and hybrid methods are useful for incorporating the shape restrictions of the density level set into the estimators. In particular, M\"uller and Sawitzki's algorithm assumes that some information about the number of clusters $M$ is given a priori. We therefore fixed two values, $M=1$ and $M=2$. Although most model densities satisfy one of these two conditions, this algorithm did not provide very competitive results. However, one of the main advantages of the excess mass methodology is that it does not need to smooth the data to reconstruct a density level set with a fixed probability content.

If, however, some geometric properties of the level set are known, then hybrid methods present a competitive alternative. For instance, if $\tau$ is small, the convex hull method was shown to provide good results. Most of these densities have convex level sets for sufficiently small values of $\tau$. Under more flexible shape restrictions, the $r$-convex hull method and granulometric smoothing could be used. However, their main disadvantage is the dependence on $r$, an unknown parameter. These approaches are very promising, because they remain quite competitive even when the value of $r$ is fixed. Selecting $r$ automatically from the sample points would significantly improve their practical performance.

}

$ $\\
\textbf{Acknowledgments.} This work has been supported by Project MTM2008-03010 of the Spanish Ministry of Science and Innovation and the IAP network StUDyS (Developing crucial Statistical methods for
Understanding major complex Dynamic Systems in natural, biomedical and social sciences) of the
Belgian Science Policy.

\bibliographystyle{plain}
\bibliography{libreria2}

\end{document}